\documentclass[twocolumn, astrosymb]{aastex701}

\usepackage{lipsum}
\usepackage{breqn}
\usepackage{amsmath}
\usepackage{gensymb}
\usepackage{hyperref}
\usepackage{ulem}
\usepackage{longtable}
\usepackage{aas_macros}
\usepackage{booktabs}
\usepackage{multirow}
\usepackage{mhchem}

\newcommand{\HII}{\ion{H}{2}~}
\newcommand{\kms}{km~s$^{-1}~$}
\newcommand{\Ha}{H$\alpha$~}
\newcommand{\HI}{\ion{H}{1}~}
\shorttitle{Chemical Signatures of Gas Accretion}
\shortauthors{Olvera et al.}
\graphicspath{{./}{Figures/}}

\begin{document}

\title{DIISC-VIII: Chemical Signatures of Gas Accretion in Star-Forming Galaxies}

\correspondingauthor{Alejandro J. Olvera}
\email{ajolver2@asu.edu}

\author[0000-0002-2819-0753, sname=Olvera, gname=Alejandro]{Alejandro J. Olvera}
\affiliation{School of Earth and Space Exploration, Arizona State University, 781 Terrace Mall, Tempe, AZ 85287, USA}
\email{ajolver2@asu.edu} 

\author[0000-0002-2724-8298, sname=Borthakur, gname=Sanchayeeta]{Sanchayeeta Borthakur}
\affiliation{School of Earth and Space Exploration, Arizona State University, 781 Terrace Mall, Tempe, AZ 85287, USA}
\email{sborthak@asu.edu}

\author[0000-0002-3472-0490, sname=Padave, gname=Mansi]{Mansi Padave}
\affiliation{Department of Astronomy $\&$ Astrophysics, University of California, San Diego, 9500 Gilman Drive, LaJolla, CA 92093, USA}
\email{mpadave@ucsd.edu}

\author[0000-0003-1436-7658, sname=Gim, gname=Hansung]{Hansung B. Gim}
\affiliation{Eureka Scientific Inc., 2452 Delmer Street Suite 100, Oakland, CA 94602, USA}
\email{cosmologisths@gmail.com}

\author[0000-0003-3168-5922, sname=Momjian, gname=Emmanuel]{Emmanuel Momjian}
\affiliation{National Radio Astronomy Observatory, P.O. Box O, Socorro, NM, USA}
\email{emomjian@nrao.edu}

\author[0000-0001-6670-6370, sname=Heckman, gname=Timothy]{Timothy M. Heckman}
\affiliation{School of Earth and Space Exploration, Arizona State University, 781 Terrace Mall, Tempe, AZ 85287, USA}
\affiliation{Department of Physics \& Astronomy, Johns Hopkins University, Baltimore, MD 21218, USA}
\email{theckma1@jhu.edu}

\begin{abstract}

We investigate anomalously low-metallicity (ALM) star-forming regions in 11 galaxies from the Deciphering the Interplay between the Interstellar Medium, Stars, and the Circumgalactic Medium (DIISC) survey. We combine optical spectroscopy of 542 H$\alpha$-bright regions from MMT/Binospec with imaging from the Vatican Advanced Technology Telescope and \ion{H}{1} 21 cm imaging from the Very Large Array. We find 75 ALM regions: 19 identified using deviations from each galaxy's radial metallicity gradient, 27 using the resolved mass--metallicity relation (rMZR), and 29 via both methods. Regions identified by both methods are metal-poor relative to their galactocentric positions and local stellar mass surface densities by $>$ 0.1 dex, making them the strongest candidates for localized metallicity dilution. We estimate accreted masses of $5.6<\log(M_{\rm acc}/{\rm M_\odot})<7.5$ for the 48 radial-gradient-selected regions, comparable to observed and simulated high-velocity cloud masses. We calculate a mean hydrogen accretion rate of $1.02\ {\rm M_\odot yr^{-1}}$ per galaxy, which is comparable to the mean star formation rate of the sample. The strongest ALM candidates lie within or near $R_{25}$, while rMZR-only regions dominate the outer population. We find no evidence that circumgalactic Ly$\alpha$ absorption along a single sightline traces disk chemical structure, as quantified by either the radial metallicity gradient or the fraction of radial-gradient-selected ALM regions. These results suggest that localized accretion of low-metallicity gas can sustain star formation in galaxies and demonstrate that ALM regions provide a spatially resolved indirect view of ongoing gas cycling.

\end{abstract}

\keywords{\uat{Galaxies}{573} --- \uat{Galaxy abundances}{574} --- \uat{Galaxy accretion}{575} --- \uat{Galaxy evolution}{594} --- \uat{HII regions}{694}}


\section{Introduction}
\label{sec:introduction}

Gas accretion replenishes the interstellar medium (ISM) consumed by star formation and is therefore central to the baryon cycle and the continued growth of galaxies \citep{Keres2005,Lilly2013,FaucherGiguereOh2023}. However, directly observing this process is challenging because the relevant gas is often diffuse and multiphase, and its inflow velocities are small compared with disk rotation, making them difficult to distinguish from warps, noncircular motions, or galactic fountains \citep{Sancisi2008,DiTeodoroPeek2021}. Absorption-line observations can sensitively detect low-column-density gas in the circumgalactic medium (CGM), but a single background sightline does not confirm that the material will reach the disk. \ion{H}{1} 21 cm emission maps trace the spatial distribution and kinematics of the neutral reservoir, yet distinguishing vertical accretion from radial transport within the disk remains problematic. These observational limitations motivate indirect approaches that search for the imprint of recently accreted gas after it begins to interact with the ISM.

\begin{figure*}[t]
    \centering
    \includegraphics[width=\linewidth]{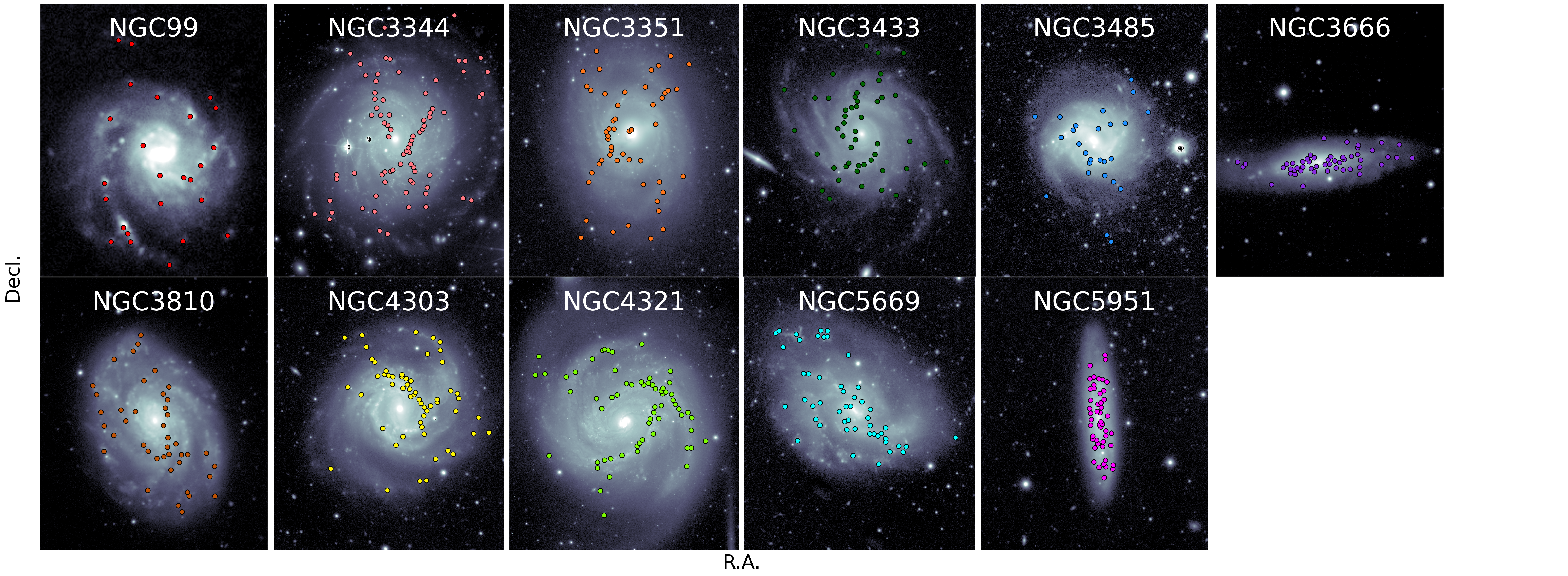}
    \caption{r-band images of 11 DIISC galaxies are shown. The circular points on each galaxy represent star-forming regions where we obtained optical spectra.}
    \label{fig:DIISC_images}
\end{figure*}

Gas-phase metallicity serves as one such imprint. Because stars produce metals and return them to the ISM, the abundance pattern across a galaxy reflects the combined effects of star formation, enrichment, inflow, outflow, and mixing \citep{FinlatorDave2008,Lilly2013}. In nearby star-forming galaxies, metallicity typically decreases with galactocentric radius \citep{Sanchez_2014A&A...563A..49S}, and local gas-phase metallicities correlate with stellar mass surface density in the resolved mass--metallicity relation (rMZR; \citealt{Yao2022A&A...661A.112Y}). When gas with a metallicity lower than that of the ambient ISM is deposited into a disk, it can locally dilute the gas before azimuthal mixing erases the contrast. Consequently, the affected star-forming regions may appear underabundant relative to the metallicity expected at their galactocentric radius or from their local stellar mass surface density.

These regions are commonly referred to as anomalously low-metallicity (ALM) regions. Using Mapping Nearby Galaxies at Apache Point Observatory (MaNGA) integral-field spectroscopy, \citet{Hwang_2019ApJ...872..144H} identified ALM spaxels as negative outliers from the rMZR and found that approximately one quarter of star-forming galaxies in their sample host a substantial ALM population. The incidence of ALM regions increases toward lower-mass galaxies and larger galactocentric radii, and these regions are often associated with young stellar populations, consistent with the expectation that metal-poor accretion both dilutes the ISM and fuels recent star formation. Studies of individual \ion{H}{2} regions provide complementary evidence. In NGC~99, \citet{Olvera_2024ApJ...976..205O} found two regions $\sim$0.16 dex below the galaxy's radial metallicity gradient and demonstrated that the accreted gas mass required to produce their metallicities is comparable to the masses of high-velocity clouds. More recently, \citet{Olvera2026arXiv260926884O} linked ALM regions in the extended disk of NGC~3344 to inward radial motions measured in \ion{H}{1}, providing direct evidence that gas transport within an extended neutral disk can dilute metallicity locally and affect subsequent star formation. Because the radial-gradient and rMZR methods define the expected metallicity relative to galactocentric radius and local stellar mass surface density, respectively, applying both may distinguish different configurations of chemically distinct gas. Regions identified by both methods may trace strong, localized dilution, while radial-gradient-only regions may reflect recent local metallicity variations and rMZR-only regions may trace gas that has mixed with the surrounding ISM but remains under-enriched relative to the local stellar population. 

In this work, we conduct a uniform search for ALM star-forming regions across 11 nearby galaxies. Section~\ref{sec:data} details the optical spectroscopy, optical imaging, and \ion{H}{1} data and their analysis. In Section~\ref{sec:results}, we identify regions that are anomalous relative to their host-galaxy radial metallicity gradient and the rMZR. Section~\ref{sec:discuss} discusses the physical mechanisms behind the two ALM diagnostics, estimates accreted gas masses and accretion rates, and explores potential connections to the CGM. Section~\ref{sec:conclusion} summarizes our findings.

\begin{deluxetable*}{lcccccccccc}[t]
\tabletypesize{\footnotesize}
\tablecaption{Properties of Galaxies in the Sample \label{tab:gal-sample}}
\tablehead{
\colhead{Galaxy} & 
\colhead{R.A.} & 
\colhead{Dec} & 
\colhead{$z$} & 
\colhead{Distance} & 
\colhead{Scale} & 
\colhead{$i$} & 
\colhead{$\rm log(M_\star/M_\odot)$} &
\colhead{SFR$_{\mathrm{UV}}$} &
\colhead{$\rm log(M_{\rm HI}/M_\odot)$} & \\
\colhead{} & 
\colhead{(deg)} & 
\colhead{(deg)} & 
\colhead{} & 
\colhead{(Mpc)} & 
\colhead{(pc arcsec$^{-1}$)} & 
\colhead{($^\circ$)} & 
\colhead{} &
\colhead{(M$_\odot$ yr$^{-1}$)} & 
\colhead{} & \\
\colhead{(1)} & 
\colhead{(2)} & 
\colhead{(3)} & 
\colhead{(4)} & 
\colhead{(5)} & 
\colhead{(6)} & 
\colhead{(7)} & 
\colhead{(8)} & 
\colhead{(9)} &
\colhead{(10)} &
}
\startdata
NGC 99   &   5.99747 & 15.77042 & 0.01772 & 79.4 & 378.17 & 20 & 10.62 & 2.45 & 10.35\\
NGC 3344 & 160.87951 & 24.92249 & 0.00196 &  8.28 &  40.14 &  6 & 10.18 & 0.47 & 9.44 \\
NGC 3351 & 160.99038 & 11.70380 & 0.00259 &  9.29 &  45.04 & 44 & 10.90 & 0.79 & 9.00\\
NGC 3433 & 163.01611 & 10.14830 & 0.00908 & 44.59 & 216.18 & 22 & 10.85 & 1.20 & 10.16\\
NGC 3485 & 165.00991 & 14.84157 & 0.00477 & 29.43 & 142.68 & 23 & 10.45 & 0.83 & 9.75\\
NGC 3666 & 171.10861 & 11.34222 & 0.00353 & 17.1  &  82.90 & 75 & 10.31 & 0.32 & 9.44\\
NGC 3810 & 175.24481 & 11.47112 & 0.00331 & 15.3  &  74.18 & 47 & 10.44 & 1.25 & 9.43\\
NGC 4303 & 185.47894 &  4.47347 & 0.00522 & 18.7  &  90.66 & 22 & 11.39 & 5.25 & 9.88\\
NGC 4321 & 185.72890 & 15.82101 & 0.00525 & 13.93 &  67.53 & 24 & 10.58 & 2.49 & 9.38\\
NGC 5669 & 218.18311 &  9.89170 & 0.00456 & 14.5  &  70.30 & 48 &  9.52 & 0.29 & 9.23\\
NGC 5951 & 233.42933 & 15.00715 & 0.00593 & 27.1  & 131.38 & 77 & 10.10 & 0.23 & 9.53\\
\enddata
\tablecomments{
Column (1): Galaxies in our sample.
Columns (2–3): J2000 celestial coordinates.
Columns (4–6): recessional velocity, adopted distance, and the physical scale at that distance in pc arcsec$^{-1}$.
Column (7): inclination estimated from the apparent axis ratio.
Column (8): stellar masses for our galaxies from \cite{Padave_2024ApJ...960...24P}
Columns (9): dust-corrected FUV SFRs.
Column (10): \ion{H}{1} masses for our galaxies from \cite{Padave_2024ApJ...960...24P}
}
\end{deluxetable*}

\section{Observations and Data Analysis} \label{sec:data}

\subsection{The xDIISC Sample}
\label{subsec:xdiisc_sample}

The 11 galaxies analyzed in this study make up the xDIISC sample, first introduced in \citet{Padave_2025ApJ...986..145P}. The xDIISC sample is derived from the Deciphering the Interplay between the Interstellar Medium, Stars, and the Circumgalactic Medium (DIISC) survey, which employs a multiwavelength approach to connect the CGM, the neutral ISM, and recent star formation within the same nearby galaxies \citep{borthakur2024diisc}. The parent DIISC galaxies were selected based on the presence of a UV-bright background quasar (QSO) at a small impact parameter relative to the \ion{H}{1} extent, facilitating ultraviolet absorption-line spectroscopy through the disk--CGM interface. Together, these multi-wavelength observations allow us to trace three stages of the proposed accretion pathway: gas in the CGM, gas in the extended \ion{H}{1} reservoir, and chemically anomalous gas within star-forming regions, all within a single, consistently analyzed sample.

The xDIISC sample spans a broad range of global properties among nearby star-forming disks: $9.52 \leq \log(M_\star/M_\odot) \leq 11.39$, $0.23 \leq {\rm SFR}/(M_\odot\,{\rm yr}^{-1}) \leq 5.25$, and $9.00 \leq \log(M_{\rm HI}/M_\odot) \leq 10.35$ \citep{Padave_2024ApJ...960...24P,Padave_2025ApJ...986..145P}. Table~\ref{tab:gal-sample} provides the global properties of each galaxy. This diversity enables the examination of ALM regions across galaxies with varying stellar masses, star formation rates, and atomic gas reservoirs. A key advantage of the xDIISC sample over those used in previous ALM studies is the availability of resolved \ion{H}{1} data for every galaxy. The \ion{H}{1} maps enable direct measurement of atomic gas mass near each \ion{H}{2} region, rather than relying on scaling relations. Since the accreted mass estimated from a dilution model depends on the pre-existing gas reservoir, these measurements offer a more robust foundation for determining accreted masses and accretion rates.

\begin{table*}[t]
\centering
\caption{VLA \HI\ Observations}
\label{tab:obs}
\begin{tabular}{lccccccc}
\hline\hline
Name & Project & Total & Flux/Bandpass & Complex Gain & Synthesized & Velocity & Sensitivity \\
 & Code & Integration & Calibrator & Calibrator & Beam & Resolution & \\
 & & (hr) & & & (arcsec $\times$ arcsec) & (km\,s$^{-1}$) & (mJy\,beam$^{-1}$ (\kms)$^{-1}$) \\
\hline
NGC\,99 & 19B-183 & 6.5 & 3C48 & J2340$+$1333 & $50.5 \times 47.5$ & 2.2 & 0.98 \\
\multirow{2}{*}{NGC\,3344} & 20A-125 & \multirow{2}{*}{52} & \multirow{2}{*}{3C286} & \multirow{2}{*}{J1021$+$2159} & \multirow{2}{*}{$6.5 \times 5.5$} & \multirow{2}{*}{3.3} & \multirow{2}{*}{0.275} \\
 & 21A-408 & & & & & & \\
NGC\,3351 & 17A-090 & 6.5 & 3C147 & J1120$+$1420 & $53 \times 51$ & 3.3 & 0.70 \\
NGC\,3433 & 17A-090 & 6.5 & 3C147 & J1120$+$1420 & $65.5 \times 51.0$ & 3.3 & 0.65 \\
\multirow{2}{*}{NGC\,3485} & 18A-006 & \multirow{2}{*}{8.5} & \multirow{2}{*}{3C286} & \multirow{2}{*}{J1120$+$1420} & \multirow{2}{*}{$54 \times 54$} & \multirow{2}{*}{3.3} & \multirow{2}{*}{0.74} \\
 & 19B-183 & & & & & & \\
\multirow{2}{*}{NGC\,3666} & 18A-006 & \multirow{2}{*}{8.7} & \multirow{2}{*}{3C286} & \multirow{2}{*}{J1120$+$1420} & \multirow{2}{*}{$60.5 \times 56.5$} & \multirow{2}{*}{3.3} & \multirow{2}{*}{0.60} \\
 & 19B-183 & & & & & & \\
\multirow{2}{*}{NGC\,3810} & 17A-090 & \multirow{2}{*}{6.5} & \multirow{2}{*}{3C286} & \multirow{2}{*}{J1120$+$1420} & \multirow{2}{*}{$39.5 \times 30.0$} & \multirow{2}{*}{3.3} & \multirow{2}{*}{0.63} \\
 & 18A-006 & & & & & & \\
NGC\,4303 & 19B-183 & 6 & 3C286 & J1150$-$0023 & $56.0 \times 48.0$ & 3.3 & 0.70 \\
NGC\,4321 & 18A-006 & 4 & 3C286 & J1254$+$1141 & $44 \times 30$ & 3.3 & 1.3 \\
NGC\,5669 & 17A-090 & 14 & 3C286 & J1432$+$0955 & $62.5 \times 53.0$ & 3.3 & 0.85 \\
NGC\,5951 & 17A-090 & 6.5 & 3C286 & J1520$+$2016 & $73.0 \times 50.5$ & 3.3 & 0.85 \\
\hline
\end{tabular}
\label{tab:vla_observations}
\end{table*}

\subsection{Optical Spectroscopy}\label{subsec:opt_spec}
We obtained multi-object optical spectroscopy of 542 \Ha-bright regions in our sample of galaxies using the Binospec spectrograph from the 6.5 m MMT Observatory located on Mt. Hopkins in Arizona, USA. The data were collected from 2020 to 2025 over a series of 14 nights. Our catalog of regions was created using \texttt{Source Extractor} \citep{Bertin_1996A&AS..117..393B} on \Ha+[\ion{N}{2}] images of our targets taken with the 1.8 m VATT. The images and selection method are presented in \cite{Padave_2024ApJ...960...24P}. The \texttt{BinoMask} software \citep{kansky_2019PASP..131g5005K} then automatically selected targets from the catalog by optimizing for the highest number of targets per mask design and placing slits to avoid effects of differential atmospheric refraction. Each mask was observed for a total exposure time of 2400 seconds, divided into four exposures of 600 seconds each. The two observations of NGC 3433 were each observed for eight exposures of 600 seconds for a total of 4800 seconds. We used Binospec’s 270 lines mm$^{-1}$ grating, centered at 5800 \AA providing a dispersion of 1.3 Å pix$^{-1}$ and the LP3500 filter. This setup enabled complete coverage of all relevant emission lines within the wavelength range of 3550–8850 \AA. The data were reduced using the Binospec Data Reduction Pipeline, which performed bias correction, flat-fielding, wavelength calibration, relative flux correction, co-addition, and extraction to 1D spectra \citep{kansky_2019PASP..131g5005K}.

The identification of emission lines and their flux measurement were carried out using the procedure described in \cite{Olvera_2024ApJ...976..205O}, with one important modification. Prior to measuring emission-line fluxes, the underlying stellar continuum of each extracted spectrum was modeled and subtracted using the Penalized Pixel-Fitting (\texttt{pPXF}) method \citep{Cappellari_2023MNRAS.526.3273C}. We used the E-MILES simple stellar population templates \citep{Vazdekis_2016MNRAS.463.3409V} distributed with \texttt{pPXF}, which are based on Padova isochrones and assume a \cite{Salpeter_1955ApJ...121..161S} initial mass function. Subtracting the best-fitting stellar continuum minimizes the effects of underlying stellar Balmer absorption, enabling more accurate measurements of the nebular emission-line fluxes. Emission-line fluxes were subsequently measured by fitting Gaussian profiles to the continuum-subtracted spectra, and only lines detected with a signal-to-noise ratio greater than three were retained for subsequent analysis.

\subsection{Optical Imaging}\label{subsec:opt_img}
We utilize stellar mass maps created by \cite{Padave_2024ApJ...960...24P} from optical continuum g- and r- band imaging taken with the Vatican Advanced Technology Telescope (VATT) to measure stellar mass surface density within $0.8\times0.8$ kpc masks centered at the position of each \HII region and integrate the pixels contained within each mask using \texttt{photutils}\citep{Bradley_larry_bradley_2023_7946442}. We choose this size to avoid spatial decorrelation between \HII regions and cold molecular gas \citep{Schinnerer_2019ApJ...887...49S,Semenov_2021ApJ...918...13S}. Additionally, we cross calibrate our H$\alpha$ fluxes by utilizing VATT narrowband images from \cite{Padave_2025ApJ...986..145P} in order to ensure an accurate SFR measurement. 

\begin{figure*}[t]
    \centering
    \includegraphics[width=\linewidth]{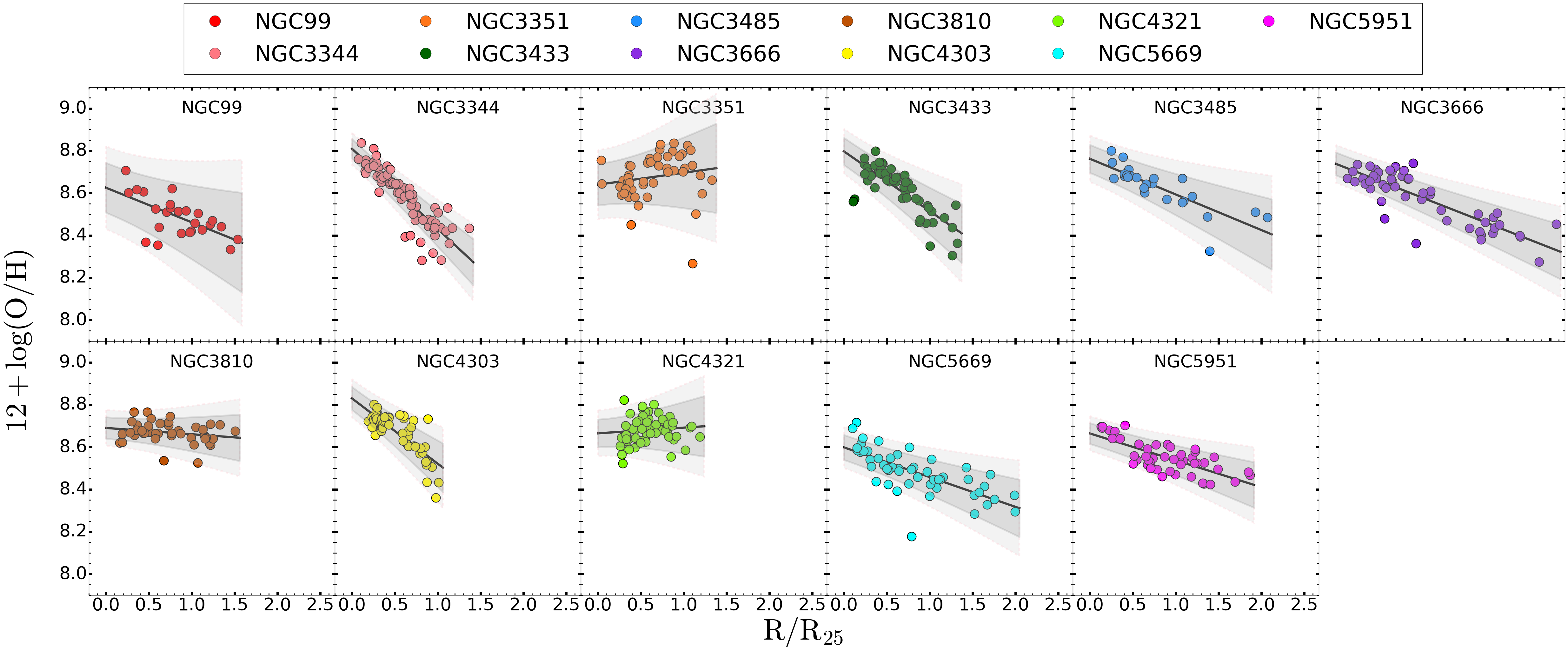}
    \caption{The radial dependence of gas-phase metallicity is shown for each of the 11 DIISC galaxies. Here, we use the N2 \cite{Curti_FMR_2020MNRAS.491..944C} strong-line calibration. Points represent the metallicity of star-forming regions in each galaxy. The solid black line in each panel shows the linear fit. The light and dark grey shaded regions represent the 3$\sigma$ and 5$\sigma$ confidence interval of each fit.}
    \label{fig:rad_grads}
\end{figure*}

\subsection{VLA HI 21cm Imaging}

\HI\ observations were obtained with NSF's Karl G. Jansky Very Large Array (VLA) under the VLA–DIISC program (P.I. S. Borthakur; project codes 17A-090, 18A-006, 19B-183, 20A-125, and 21A-408). The majority of the sample was observed in the D-configuration (maximum baseline $B_{\rm max}\sim$ 1 km), supplemented by follow-up B- and C-configuration observations ($B_{\rm max}\sim$ 11 km and 3 km, respectively) for a subset of targets. The data were calibrated in the Common Astronomy Software Applications package \citep[CASA; ][]{CASA_2022PASP..134k4501C} using a semi-automatic pipeline that solved for delay, bandpass, and complex gain (amplitude and phase) terms. Hanning smoothing was applied to suppress Gibbs ringing, degrading the spectral resolution to twice the native channel width and yielding final velocity resolutions of 2.2 \kms\ for NGC 99 and 3.3 \kms\ for the remaining targets. Visibilities affected by radio-frequency interference were flagged. Table~\ref{tab:vla_observations} summarizes the observations for each target, listing the project code, total integration time, flux density scale/bandpass and phase calibrators, synthesized beam size, velocity resolution, and image-cube sensitivity.

Table~\ref{tab:vla_observations} is identical to Table~2 of \citet{Padave_2024ApJ...960...24P} except for NGC~3344, for which \citet{Padave_2024ApJ...960...24P} used archival D-configuration data at a coarser velocity resolution of 10.36 \kms. The cube presented here instead incorporates new D-configuration observations (project code 21A-408) matched in velocity resolution to the B- and C-configuration data (project code 20A-125), and therefore supersedes the earlier one.

Continuum emission was subtracted in the visibility domain with the CASA task uvcontsub, and the resulting spectral-line visibilities were imaged with tclean using the Clark deconvolution algorithm and the barycentric spectral reference frame. Cubes were produced with Briggs weighting and a robust parameter of 0.5, which balances point-source sensitivity against synthesized beam size. NGC 4321 is the exception, having been imaged with uniform weighting; see \citet{Gim_2021ApJ...922...69G} for details.

\begin{figure}[h]
    \centering
    \includegraphics[width=\linewidth]{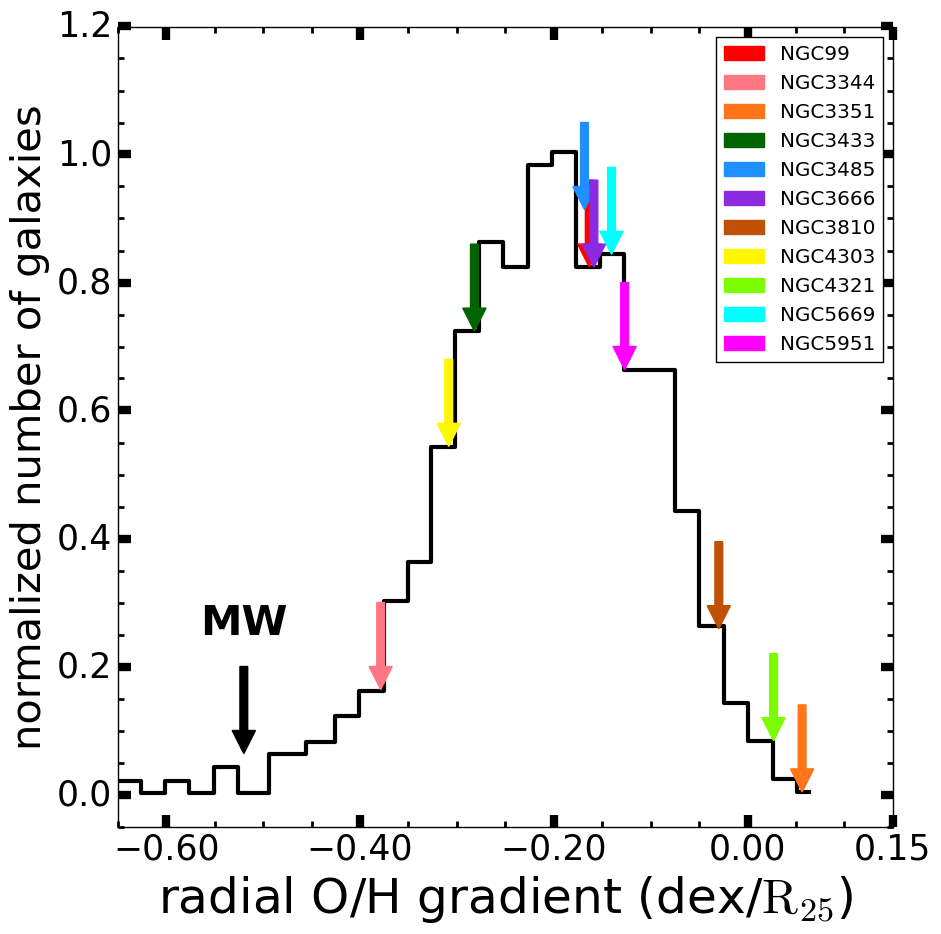}
    \caption{The metallicity gradients in units of dex R$_{25}^{-1}$ of the 11 DIISC galaxies are shown in comparison to the gradients of a larger sample of galaxies from the MaNGA survey. Each colored arrow corresponds to one galaxy's gradient. The solid black histogram is the distribution from the MaNGA survey. The solid black arrow represents the metallicity gradient of the Milky Way.}
    \label{fig:metal_grad_hist}
\end{figure}

\section{Results} \label{sec:results}
\subsection{Radial Gradients}\label{subsec:RGs}
We utilize the N2 metallicity indicator, defined to be ${\rm log}($[\ion{N}{2}]$\lambda6583/{\rm H}\alpha)$, and convert to gas-phase metallicities using the strong-line calibration from \cite{Curti_FMR_2020MNRAS.491..944C}. While other optical strong-lines, such as [\ion{O}{3}] and [\ion{S}{2}], are present in many of our spectra, we choose the N2 metallicity indicator for the following reasons: (1) dust attenuation effects are minimized due to the small wavelength separation between the H$\alpha$ and [\ion{N}{2}] emission lines (2) N2 has no turnover at low or high metallicity (3) the H$\alpha$ and [\ion{N}{2}] emission lines are present in all of our spectra yielding the highest number of regions per galaxy, which increases our probability of finding anomalously low metallicity regions. Further discussion of other metallicity indicators is given in Appendix \ref{appx:AltIndicators}.

\begin{figure*}[t]
    \centering
    \includegraphics[width=\linewidth]{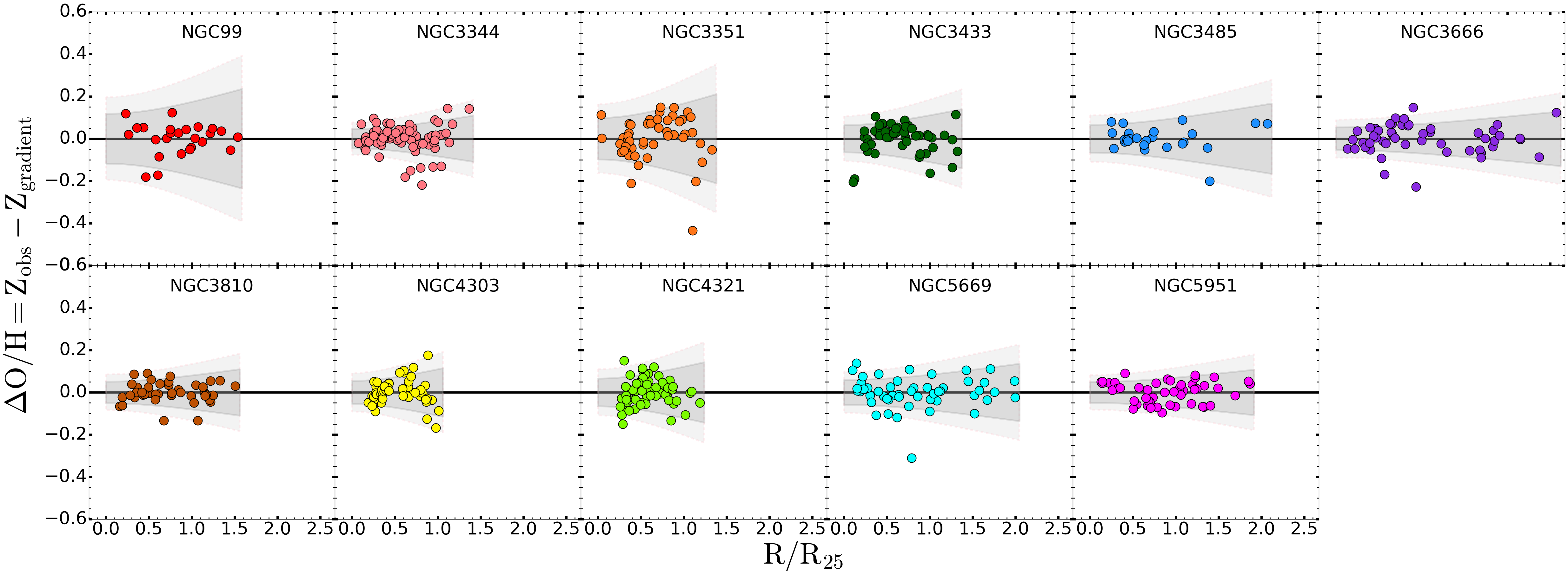}
    \caption{Same as Figure \ref{fig:rad_grads}, the y-axis is the metallicity deviation $\rm \Delta O/H \equiv Z_{obs}-Z_{gradient}$. The solid black line represents zero metallicity deviation.}
    \label{fig:Z_residuals}
\end{figure*}

In Figure \ref{fig:rad_grads}, we show each galaxy's gas-phase metallicity as a function of normalized galactocentric radius. The metallicity gradient of each galaxy is found using \texttt{lmfit} \citep{Newville_2021zndo....598352N}. Table \ref{tab:metallicity_gradients} gives each galaxy's metallicity gradient in both the natural and normalized units. The dark gray and light gray regions indicate the 3$\sigma$ and 5$\sigma$ confidence intervals of the fitted radial gradient, where $\sigma$ is the scatter about the fit. Many of galaxies in this sample have a negative metallicity gradient, which is characteristic of most normal star-forming galaxies in the local Universe \citep{Zaritsky_1994ApJ...420...87Z, Sanchez_2014A&A...563A..49S, Ho_2015MNRAS.448.2030H,Sanchez-Menguiano_2016A&A...587A..70S,Belfiore_2017MNRAS.469..151B}. Interestingly, NGC~3351 and NGC~4321 exhibit slightly positive metallicity gradients while NGC~3810 shows a very shallow gradient, hinting at the potential influence of past tidal interactions \citep{Kewley_2010ApJ...721L..48K, Gim_2021ApJ...922...69G}, radial gas flows due to a bar \cite{Martin_1994ApJ...424..599M, Sanchez-Menguiano_2026MNRAS.545f2029S}, or gas redistribution due to feedback-driven turbulence or a galactic fountain \citep{Werk_2011ApJ...735...71W, Bresolin_2012ApJ...750..122B}. 

Figure \ref{fig:metal_grad_hist} shows how the metallicity gradients of our galaxies compare to a larger population of star-forming galaxies observed as a part of of the Mapping Nearby
Galaxies at Apache Point Observatory (MaNGA) survey \citep{Pilyugin_2023A&A...676A..57P}.
Overall, the metallicity gradients of the DIISC galaxies lie within expectation in comparison to the larger sample.

\begin{table*}[!t]
\centering
\caption{Radial metallicity gradient fits for the DIISC galaxies.}
\label{tab:metallicity_gradients}
\begin{tabular}{lccccc}
\hline
Galaxy Name&
$R_{25}$ &
$N$ &
Slope &
Slope &
Intercept \\
&
(kpc) &
&
(dex kpc$^{-1}$) &
(dex $R_{25}^{-1}$) &
\\
\hline
NGC 99   & 11.46 & 25 & $-0.014 \pm 0.004$ & $-0.16 \pm 0.04$ & $8.63 \pm 0.04$ \\
NGC 3344 &  7.93 & 76 & $-0.048 \pm 0.003$ & $-0.38 \pm 0.02$ & $8.81 \pm 0.02$ \\
NGC 3351 &  8.15 & 51 & $\phantom{-}0.007 \pm 0.006$ & $\phantom{-}0.06 \pm 0.05$ & $8.64 \pm 0.03$ \\
NGC 3433 & 17.66 & 52 & $-0.016 \pm 0.002$ & $-0.28 \pm 0.03$ & $8.80 \pm 0.02$ \\
NGC 3485 &  9.87 & 26 & $-0.017 \pm 0.002$ & $-0.17 \pm 0.02$ & $8.76 \pm 0.02$ \\
NGC 3666 &  4.90 & 51 & $-0.032 \pm 0.003$ & $-0.16 \pm 0.02$ & $8.74 \pm 0.02$ \\
NGC 3810 &  7.54 & 42 & $-0.004 \pm 0.003$ & $-0.03 \pm 0.02$ & $8.69 \pm 0.02$ \\
NGC 4303 & 17.37 & 57 & $-0.018 \pm 0.002$ & $-0.30 \pm 0.03$ & $8.83 \pm 0.02$ \\
NGC 4321 & 12.85 & 63 & $\phantom{-}0.002 \pm 0.003$ & $\phantom{-}0.03 \pm 0.03$ & $8.66 \pm 0.02$ \\
NGC 5669 &  6.52 & 50 & $-0.022 \pm 0.003$ & $-0.14 \pm 0.02$ & $8.60 \pm 0.02$ \\
NGC 5951 &  6.88 & 49 & $-0.018 \pm 0.002$ & $-0.13 \pm 0.02$ & $8.66 \pm 0.02$ \\
\hline
\end{tabular}
\tablecomments{$\rm R_{25}$ values are taken from \cite{Padave_2025ApJ...986..145P}. The number of regions used in each fit N is given in the third column. }
\end{table*}

\subsection{ALM Identification}

Given the sample of star-forming regions we have and their corresponding metallicities, we aim to build a sample of ALM regions in order to investigate their characteristics. Regions can be identified as anomalously low in metallicity relative to: (1) the expected metallicity for a given radius (2) the expected metallicity for a given stellar mass and (3) a kinematic difference between the star-forming and ISM gas. In the next two subsections, we identify ALM regions using the metallicity gradients presented in Section \ref{subsec:RGs} as well as the Mass-Metallicity Relation (MZR). The third option requires having high spatial resolution ($\sim$1$^{\prime\prime}$) \ion{H}{1} data in order to accurately measure a localized \ion{H}{1} velocity and compare to high spatial resolution H$\alpha$ velocities from spectroscopy. Such data are not readily available for this sample, so we only use the first two methods.

\subsubsection{via Metallicity Gradients}\label{subsubsec:met_grad}

To identify ALMs in our sample, we define the metallicity deviation as $\rm \Delta O/H \equiv Z_{obs} - Z_{gradient} = (12+log[O/H])_{obs} - (12+log[O/H])_{gradient}$ and look for regions with  $\rm \Delta O/H < -3\sigma$. Figure \ref{fig:Z_residuals} shows $\rm \Delta O/H$ of each region as a function of normalized radius for each galaxy. Overall, a majority of regions in each galaxy fall between $3\sigma$ and $-3\sigma$. To quantify the number of ALM regions in each galaxy using the metallicity gradient, we define three classification for the regions. Regions with $\rm \Delta O/H > -3$$\sigma$ are classified as non-ALM. Conversely, regions with $-3\sigma \geq \rm \Delta O/H \geq -5$$\sigma$ are called ALM candidates while regions with $\rm \Delta O/H < -5$$\sigma$ are called robust ALM regions. Using this classification, we identify 30 ALM candidates and 18 robust ALM regions using the radial gradient method. Table \ref{tab:alm_regions} shows which regions were identified as a candidate or robust ALM region.

\subsubsection{via the Mass-Metallicity Relation}
\begin{figure}[!t]
    \centering
    \includegraphics[width=\linewidth]{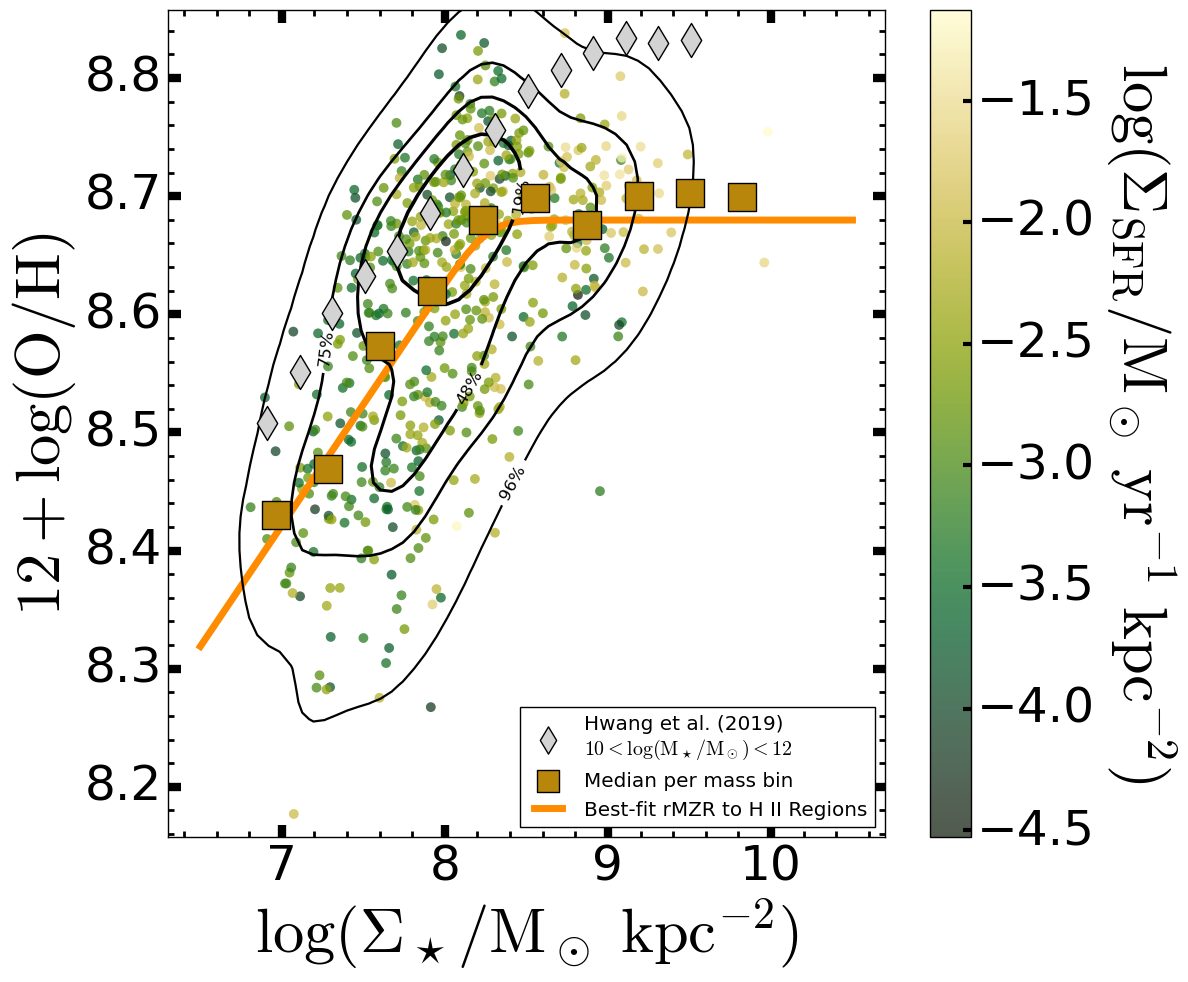}
    \caption{The resolved Mass-Metallicity Relation for the regions within the DIISC galaxies is shown, with the circular colored points representing each region. Points are color-coded by their SFR surface density. The solid black contours encapsulate 19\%, 48\%, 75\%, and 96\% percent of the points. The large brown square-shaped points represent the median metallicity per mass bin while the solid orange line is the fitted rMZR using the function form from \cite{Curti_FMR_2020MNRAS.491..944C}. The light gray diamond points represent the rMZR found by \cite{Hwang_2019ApJ...872..144H} for galaxies with a global stellar mass between $\rm 10 < log(M_\star/M_\odot)<12$.}
    \label{fig:MZR}
\end{figure}

The rMZR allows for local gas-phase metallicity to be estimated by stellar mass surface density. Since low metallicity gas accretion would cause the rMZR to overestimate the metallicity of each region if the gas has not had enough time to mix with its surroundings, we can use the relation as a second ALM region identification method. Figure \ref{fig:MZR} shows the rMZR for 541 regions. Here, the sample decreases by one due to an unreliable stellar mass estimate from the $g-r$ imaging. We fit our data with the functional form of the MZR from \cite{Curti_FMR_2020MNRAS.491..944C} given by

\begin{equation}\label{equ:MZR}
    {\rm 12+log(O/H) = Z_0} - \frac{\gamma}{\beta}*{\rm log \Bigl(1+\Bigl(\frac{\Sigma_\star}{\Sigma_0}\Bigr)}^{-\beta}\Bigr).
\end{equation}

$\rm Z_0$ and $\Sigma_0$ describe the metallicity and stellar mass surface density, respectively, at which the relation plateaus. When the stellar mass is less than $\Sigma_0$, the rMZR becomes a power law of index $\gamma$. The $\beta$ parameter controls the width of the transition region between the power law and flat sections the rMZR. Fitting our data, we find the following values: $\rm Z_0 = 8.67$ dex, ${\rm log}(\Sigma_0) = 8.26$, $\gamma = 0.20$, and $\beta=5.68$. The fit is shown by the solid orange line in Figure \ref{fig:MZR}. 

In order to investigate whether our data truly follows the functional form of the rMZR rather than a non-parametrized relation, we bin our data by mass and take the median within each bin. The large brown square-shaped points show the median metallicity in each mass bin. The median points follow the functional form of the rMZR closely demonstrating that the shape and form of the global MZR persists on smaller scale. 

For comparison, we plot the rMZR found by \cite{Hwang_2019ApJ...872..144H} as light gray diamond points for galaxies with global stellar masses between $10^{10}~{\rm M_\odot}$ and $10^{12}~{\rm M_\odot}$, which best matches our sample. The low stellar surface density slope is similar in both relations, but the flattening metallicity differs by about 0.15 dex. When $\rm log(\Sigma_\star) \leq 8.5$, the offset is about 0.1 dex. Different metallicity indicators have been shown to differ in absolute metallicity by 0.1 to 0.7 dex \citep{Kewley_2008ApJ...681.1183K}. Since we use the N2 indicator while \cite{Hwang_2019ApJ...872..144H} uses O2N2 ([\ion{O}{2}]3927,9/[\ion{N}{2}]6583), an offset in absolute metallicity calibration between the two rMZRs is expected.

\begin{longtable}[th]{lccc cc cc l}
\caption{Candidate and robust anomalously low metallicity (ALM) H\,{\sc ii} regions identified using the radial metallicity gradient and resolved mass metallicity relation (rMZR) methods. A check mark indicates that a region satisfies the corresponding ALM criterion.}
\label{tab:alm_regions}\\

\toprule
Galaxy &
$R$ &
R.A. &
Decl. &
\multicolumn{2}{c}{Radial Gradient Method} &
\multicolumn{2}{c}{rMZR Method} &
Identification Method \\
\cmidrule(lr){5-6}
\cmidrule(lr){7-8}
&
(kpc) &
(deg) &
(deg) &
Candidate &
Robust &
Candidate &
Robust &
\\
\midrule
\endfirsthead

\multicolumn{9}{c}%
{{\tablename\ \thetable{} -- continued from previous page}} \\
\toprule
Galaxy &
$R$ &
R.A. &
Decl. &
\multicolumn{2}{c}{Radial Gradient Method} &
\multicolumn{2}{c}{rMZR Method} &
ALM Method \\
\cmidrule(lr){5-6}
\cmidrule(lr){7-8}
&
(kpc) &
(deg) &
(deg) &
Candidate &
Robust &
Candidate &
Robust &
\\
\midrule
\endhead

\midrule
\multicolumn{9}{r}{{Continued on next page}} \\
\endfoot

\bottomrule
\endlastfoot

NGC 99   & 7.1  & 5.9922   & 15.7714 & ---          & ---          & $\checkmark$ & ---          & rMZR only \\
NGC 99   & 5.3  & 5.9935   & 15.7696 & $\checkmark$ & ---          & ---          & $\checkmark$ & Both \\
NGC 99   & 6.9  & 5.9946   & 15.7744 & $\checkmark$ & ---          & ---          & $\checkmark$ & Both \\
NGC 99   & 16.6 & 6.0007   & 15.7816 & ---          & ---          & ---          & $\checkmark$ & rMZR only \\
NGC 99   & 11.2 & 6.0008   & 15.7776 & ---          & ---          & ---          & $\checkmark$ & rMZR only \\
NGC 99   & 11.4 & 6.0014   & 15.7634 & ---          & ---          & ---          & $\checkmark$ & rMZR only \\

\addlinespace
NGC 3344 & 8.3  & 160.8364 & 24.9646 & $\checkmark$ & ---          & $\checkmark$ & ---          & Both \\
NGC 3344 & 4.9  & 160.8620 & 24.8924 & ---          & $\checkmark$ & $\checkmark$ & ---          & Both \\
NGC 3344 & 2.5  & 160.8621 & 24.9296 & ---          & $\checkmark$ & ---          & ---          & Radial gradient only \\
NGC 3344 & 1.3  & 160.8699 & 24.9218 & $\checkmark$ & ---          & ---          & ---          & Radial gradient only \\
NGC 3344 & 1.2  & 160.8718 & 24.9179 & $\checkmark$ & ---          & ---          & ---          & Radial gradient only \\
NGC 3344 & 6.4  & 160.8819 & 24.9665 & ---          & $\checkmark$ & ---          & ---          & Radial gradient only \\
NGC 3344 & 6.5  & 160.8843 & 24.9672 & ---          & $\checkmark$ & $\checkmark$ & ---          & Both \\
NGC 3344 & 7.5  & 160.8857 & 24.8708 & ---          & $\checkmark$ & ---          & $\checkmark$ & Both \\
NGC 3344 & 5.4  & 160.9155 & 24.9038 & ---          & $\checkmark$ & $\checkmark$ & ---          & Both \\
NGC 3344 & 8.7  & 160.9294 & 24.8826 & ---          & ---          & $\checkmark$ & ---          & rMZR only \\

\addlinespace
NGC 3351 & 9.0 & 160.9592 & 11.7421 & ---          & $\checkmark$ & ---          & $\checkmark$ & Both \\
NGC 3351 & 3.8 & 160.9948 & 11.7261 & $\checkmark$ & ---          & $\checkmark$ & ---          & Both \\
NGC 3351 & 3.2 & 161.0046 & 11.7042 & ---          & $\checkmark$ & ---          & $\checkmark$ & Both \\
NGC 3351 & 9.3 & 161.0147 & 11.6552 & $\checkmark$ & ---          & $\checkmark$ & ---          & Both \\

\addlinespace
NGC 3433 & 22.3 & 163.0152 & 10.1759 & $\checkmark$ & ---          & ---          & $\checkmark$ & Both \\
NGC 3433 & 2.1  & 163.0182 & 10.1465 & ---          & $\checkmark$ & ---          & ---          & Radial gradient only \\
NGC 3433 & 1.9  & 163.0183 & 10.1491 & ---          & $\checkmark$ & ---          & ---          & Radial gradient only \\
NGC 3433 & 17.7 & 163.0260 & 10.1279 & $\checkmark$ & ---          & ---          & $\checkmark$ & Both \\

\addlinespace
NGC 3485 & 13.8 & 164.9969 & 14.8609 & --- & $\checkmark$ & $\checkmark$ & --- & Both \\

\addlinespace
NGC 3666 & 10.5 & 171.0782 & 11.3445 & ---          & ---          & $\checkmark$ & ---          & rMZR only \\
NGC 3666 & 8.3  & 171.0862 & 11.3447 & ---          & ---          & $\checkmark$ & ---          & rMZR only \\
NGC 3666 & 11.6 & 171.0882 & 11.3492 & ---          & ---          & ---          & $\checkmark$ & rMZR only \\
NGC 3666 & 8.5  & 171.0912 & 11.3467 & ---          & ---          & $\checkmark$ & ---          & rMZR only \\
NGC 3666 & 4.6  & 171.0952 & 11.3391 & ---          & $\checkmark$ & ---          & $\checkmark$ & Both \\
NGC 3666 & 6.0  & 171.0960 & 11.3453 & ---          & ---          & $\checkmark$ & ---          & rMZR only \\
NGC 3666 & 8.1  & 171.0960 & 11.3475 & ---          & ---          & $\checkmark$ & ---          & rMZR only \\
NGC 3666 & 2.8  & 171.1006 & 11.3403 & ---          & $\checkmark$ & $\checkmark$ & ---          & Both \\
NGC 3666 & 2.6  & 171.1057 & 11.3398 & $\checkmark$ & ---          & ---          & ---          & Radial gradient only \\
NGC 3666 & 9.3  & 171.1071 & 11.3503 & ---          & ---          & $\checkmark$ & ---          & rMZR only \\
NGC 3666 & 9.0  & 171.1136 & 11.3349 & ---          & ---          & $\checkmark$ & ---          & rMZR only \\
NGC 3666 & 10.5 & 171.1238 & 11.3351 & ---          & ---          & $\checkmark$ & ---          & rMZR only \\
NGC 3666 & 7.6  & 171.1324 & 11.3416 & ---          & ---          & $\checkmark$ & ---          & rMZR only \\
NGC 3666 & 8.3  & 171.1332 & 11.3407 & ---          & ---          & $\checkmark$ & ---          & rMZR only \\
NGC 3666 & 8.2  & 171.1350 & 11.3422 & ---          & ---          & $\checkmark$ & ---          & rMZR only \\

\addlinespace
NGC 3810 & 5.1 & 175.2364 & 11.4601 & ---          & $\checkmark$ & $\checkmark$ & --- & Both \\
NGC 3810 & 1.4 & 175.2406 & 11.4730 & $\checkmark$ & ---          & ---          & --- & Radial gradient only \\
NGC 3810 & 1.2 & 175.2420 & 11.4696 & $\checkmark$ & ---          & ---          & --- & Radial gradient only \\
NGC 3810 & 8.1 & 175.2497 & 11.4958 & $\checkmark$ & ---          & ---          & --- & Radial gradient only \\

\addlinespace
NGC 4303 & 17.0 & 185.4296 & 4.4618 & $\checkmark$ & --- & ---          & $\checkmark$ & Both \\
NGC 4303 & 15.2 & 185.4575 & 4.5110 & $\checkmark$ & --- & $\checkmark$ & ---          & Both \\
NGC 4303 & 4.7  & 185.4653 & 4.4750 & $\checkmark$ & --- & ---          & ---          & Radial gradient only \\
NGC 4303 & 4.1  & 185.4773 & 4.4852 & $\checkmark$ & --- & ---          & ---          & Radial gradient only \\
NGC 4303 & 17.5 & 185.5159 & 4.4404 & ---          & --- & ---          & $\checkmark$ & rMZR only \\

\addlinespace
NGC 4321 & 4.7  & 185.7112 & 15.8303 & $\checkmark$ & ---          & ---          & --- & Radial gradient only \\
NGC 4321 & 3.6  & 185.7188 & 15.8125 & $\checkmark$ & ---          & ---          & --- & Radial gradient only \\
NGC 4321 & 3.7  & 185.7205 & 15.8106 & $\checkmark$ & ---          & ---          & --- & Radial gradient only \\
NGC 4321 & 11.0 & 185.7370 & 15.8620 & $\checkmark$ & ---          & ---          & --- & Radial gradient only \\
NGC 4321 & 5.5  & 185.7370 & 15.8026 & $\checkmark$ & ---          & ---          & --- & Radial gradient only \\
NGC 4321 & 3.7  & 185.7414 & 15.8301 & ---          & $\checkmark$ & $\checkmark$ & --- & Both \\

\addlinespace
NGC 5669 & 10.9 & 218.1504 & 9.8823 & ---          & ---          & $\checkmark$ & ---          & rMZR only \\
NGC 5669 & 6.5  & 218.1718 & 9.8783 & $\checkmark$ & ---          & ---          & ---          & Radial gradient only \\
NGC 5669 & 2.4  & 218.1814 & 9.8991 & ---          & $\checkmark$ & $\checkmark$ & ---          & Both \\
NGC 5669 & 9.9  & 218.1910 & 9.9174 & ---          & ---          & $\checkmark$ & ---          & rMZR only \\
NGC 5669 & 3.4  & 218.1955 & 9.8892 & $\checkmark$ & ---          & ---          & ---          & Radial gradient only \\
NGC 5669 & 4.0  & 218.1964 & 9.8934 & ---          & $\checkmark$ & $\checkmark$ & ---          & Both \\
NGC 5669 & 5.2  & 218.2016 & 9.8825 & ---          & $\checkmark$ & ---          & $\checkmark$ & Both \\
NGC 5669 & 13.0 & 218.2078 & 9.9170 & ---          & ---          & $\checkmark$ & ---          & rMZR only \\

\addlinespace
NGC 5951 & 9.1 & 233.4257 & 14.9986 & ---          & --- & $\checkmark$ & --- & rMZR only \\
NGC 5951 & 4.9 & 233.4268 & 15.0080 & $\checkmark$ & --- & $\checkmark$ & --- & Both \\
NGC 5951 & 4.7 & 233.4274 & 15.0033 & ---          & --- & $\checkmark$ & --- & rMZR only \\
NGC 5951 & 8.1 & 233.4281 & 14.9927 & ---          & --- & $\checkmark$ & --- & rMZR only \\
NGC 5951 & 5.0 & 233.4285 & 14.9983 & ---          & --- & $\checkmark$ & --- & rMZR only \\
NGC 5951 & 3.8 & 233.4300 & 14.9992 & $\checkmark$ & --- & $\checkmark$ & --- & Both \\
NGC 5951 & 3.5 & 233.4305 & 15.0003 & $\checkmark$ & --- & $\checkmark$ & --- & Both \\
NGC 5951 & 6.7 & 233.4311 & 15.0169 & ---          & --- & $\checkmark$ & --- & rMZR only \\
NGC 5951 & 4.6 & 233.4317 & 15.0016 & $\checkmark$ & --- & $\checkmark$ & --- & Both \\
NGC 5951 & 5.4 & 233.4321 & 15.0070 & $\checkmark$ & --- & $\checkmark$ & --- & Both \\
NGC 5951 & 7.3 & 233.4323 & 15.0131 & ---          & --- & $\checkmark$ & --- & rMZR only \\
NGC 5951 & 5.8 & 233.4324 & 15.0051 & $\checkmark$ & --- & $\checkmark$ & --- & Both \\

\end{longtable}

\begin{figure}[h]
    \centering
    \includegraphics[width=\linewidth]{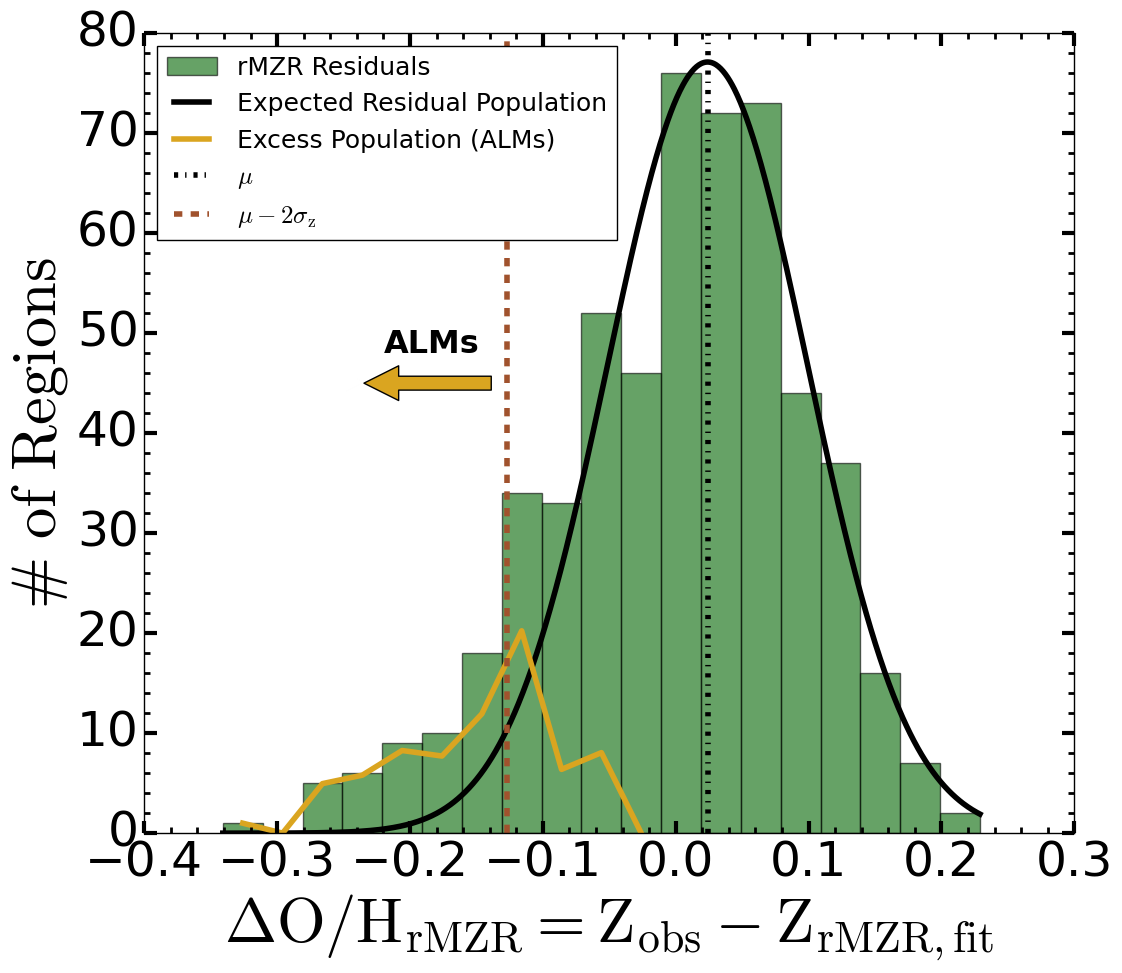}
    \caption{The rMZR metallicity deviation distribution for the regions within the 11 DIISC galaxies is shown by the green histogram. The solid black curve represents a Gaussian fit to the positive side of the histogram, which is then assumed to be symmetric. The dash-dot vertical line shows the center of the distribution. The brown dashed line indicates the metallicity deviation which is -2$\sigma_z$ away from the center. The gold solid line represents the number of ALM regions found by subtracting the symmetric Gaussian distribution from the rMZR residual distribution from}
    \label{fig:MZR_residual}
\end{figure}

Since the median points closely follow the functional rMZR fit, we take the latter to be our expected metallicity $\rm Z_{MZR,fit}$. We then take the difference between the observed metallicity and $\rm Z_{MZR,fit}$ to be the rMZR metallicity deviation. Figure \ref{fig:MZR_residual} shows the rMZR metallicity deviation distribution of the DIISC regions. The positive side of the distribution appears normal; however, the negative side has a low metallicity tail. A small number regions would be expected to show larger negative metallicity deviation, assuming a Gaussian distribution, but we observe more regions than expected. A similar $\Delta \rm O/H_{rMZR}$ distribution shape is seen by \cite{Hwang_2019ApJ...872..144H}. To estimate the excess number of regions, we fit a Gaussian to the positive side of the distribution and, then, assume it to be symmetric, which results in the black curve shown. We find the center of the distribution $\mu$ to be 0.0240 dex and the standard deviation $\sigma_{\rm z}$ to be 0.07 dex. Our standard deviation value is the same as that found by \cite{Hwang_2019ApJ...872..144H}. Following their methodology to quantify the ALM regions, we subtract the fitted Gaussian from the rMZR residuals distribution. The result is the gold solid line, which reveals the excess population of regions hidden within the original distribution. The excess population shows larger metallicity deviation than the rest of the distribution meaning that the observed metallicity is overestimated by the rMZR. Low metallicity gas accretion is likely diluting the existing \HII regions to levels well below expectation, which causes the metallicity deviation to increase. Following \cite{Hwang_2019ApJ...872..144H}, we define the ALM \HII regions to be any region with $\Delta {\rm O/H_{rMZR}} < \mu-2\sigma_{\rm z}$, which is shown by the brown dashed line and gold left-pointing arrow. In total, we identify 56 ALM \HII regions.

As in Section \ref{subsubsec:met_grad}, we classify regions into ALM candidates and robust ALMs using the standard deviation of the rMZR. ALMs candidates are those with $\mu-3\sigma_{\rm z} < \Delta {\rm O/H_{rMZR}} < \mu-2\sigma_{\rm z}$ while robust ALMs are those with $\Delta {\rm O/H_{rMZR}} < \mu-3\sigma_{\rm z}$. Table \ref{tab:alm_regions} gives which regions have been identified as robust or candidate.

\section{Discussion} \label{sec:discuss}
\subsection{What the Two Methods Identify}\label{subsec:method_comparison}

We identify ALM regions using two methods that compare the measured gas-phase metallicity to different expected values. The radial-gradient method determines whether a region is metal-poor relative to other regions at a similar galactocentric radius within the same galaxy. The resolved mass metallicity relation (rMZR) method instead compares the measured metallicity with that expected from the region's stellar mass surface density. The two methods identify overlapping but different ALM populations. Of the 75 total ALM regions, 19 are identified only by the radial-gradient method, 27 are identified only by the rMZR, and 29 are identified by both methods.

Regions identified by both methods are our strongest candidates for recently diluted gas because they are metal-poor relative to both their local surroundings and their stellar mass surface densities. The accretion of low metallicity gas can lower the metallicity of the pre-existing ISM while increasing the amount of gas available for star formation. Previous observations have also found that ALM regions are preferentially associated with elevated local and global specific star formation rates, supporting a connection between gas accretion, metallicity dilution, and enhanced star formation \citep{Hwang_2019ApJ...872..144H}.
\subsection{Physical Interpretation of the ALM Classifications}\label{subsec:alm_interpretation}

We propose that ALM regions identified using the radial-gradient and rMZR methods trace different configurations of gas flows in and around galactic disks. To build this physical interpretation, we use the local orbital period, $t_{\rm orb}$, as a temporal signpost. The orbital period is the time required for material to complete one orbit around the galactic center and is given by $t_{\rm orb}=2\pi R / V_{\rm circ}$, where $V_{\rm circ}$ is the circular velocity. The orbital period of the stars in the solar neighborhood is approximately 200–250 Myr, while orbital periods across extended galactic disks can range from a few hundred Myr to approximately 1 Gyr \citep{Wong_2009ApJ...705..650W}. Simulations find that randomly distributed metals can become homogenized on timescales comparable to or shorter than the local orbital period \citep{YangKrumholz2012}. In particular, differential rotation winds non-axisymmetric chemical structure into progressively smaller spatial scales, where turbulence can erase the initial metallicity contrast in less than one orbit \citep{Petit2015}. We therefore expect the chemical signature of a localized gas-flow event to become less distinct from the surrounding ISM over a timescale $t_{\rm mix}\lesssim t_{\rm orb}$.

The simplest case may be ALM regions identified only by the radial-gradient method. These regions have lower metallicities than other regions at a similar galactocentric radius but remain consistent with the metallicity expected from their local stellar mass surface densities. This configuration can result from the localized vertical accretion of low-metallicity gas onto the disk. Possible sources include HVCs and IVCs similar to those observed around the Milky Way, recycled gas that has mixed with lower-metallicity material in the CGM, and/or gas accreted from a low-mass satellite \citep{Putman_2012ARA&A..50..491P,Wakker1999LowMetallicityAccretion,Wakker2001DistancesMetallicities,Bregman1980GalacticFountain,FraternaliBinney2008Accretion,TothOstriker1992DiskInfall}. The kinematically settled appearance of the H I disks in our sample makes an ongoing disruptive interaction an unlikely explanation for most of these regions, although past or low-mass accretion events cannot be excluded.

A localized metallicity deficit should remain detectable relative to the radial gradient only while the accreted material is chemically distinct from its surroundings. Radial-gradient-only regions may therefore trace relatively recent dilution events for which $t\lesssim t_{\rm mix}$. Robust ALM regions may represent gas that has undergone little mixing with the surrounding ISM, corresponding approximately to $t\ll t_{\rm mix}$. Candidate ALM regions may instead trace weaker events, lower accreted gas masses, or gas for which mixing has already reduced the initial metallicity contrast. These possibilities cannot be distinguished from metallicity alone, but they provide a physical interpretation for localized deviations from a galaxy’s radial metallicity distribution.

ALM regions identified only by the rMZR may represent a different chemical state of the disk. The rMZR relates the present gas-phase metallicity to the local stellar mass surface density, which records the cumulative conversion of gas into stars and the associated enrichment history. A region with a higher stellar mass surface density than other regions at the same galactocentric radius may therefore have a higher expected metallicity from the rMZR than from the radial gradient. If the gas initially reflects this enhanced chemical evolution, it can appear metal-rich relative to the radial gradient while remaining consistent with the rMZR.

Turbulence and differential rotation can subsequently redistribute this enriched gas and mix it with the surrounding ISM. Mixing drives the gas-phase metallicity toward the mean abundance at that galactocentric radius, while the underlying stellar mass surface density remains comparatively unchanged. The gas may therefore become consistent with the radial gradient but remain underabundant relative to the rMZR. In this scenario, rMZR-only ALM regions trace locations where the present gas metallicity has become decoupled from the previous enrichment history recorded by the stellar population. The relevant condition is $t\gtrsim t_{\rm mix}$. The underlying stellar structure persists while successive generations of H II regions illuminate gas that is exchanged and mixed within the disk.

Regions identified by both methods may represent the clearest cases of strong, localized dilution. Their metallicities are low relative to both the surrounding gas at the same galactocentric radius and the chemical-enrichment history implied by the local stellar mass surface density. This can occur when the mass of accreted gas is large enough, or its metallicity low enough, to lower the abundance below both expectations. In this interpretation, the dilution occurred recently enough that mixing has not erased the contrast with the surrounding ISM, while its strength is sufficient to produce an rMZR offset. Regions identified by both methods may therefore trace strong accretion events observed at $t\lesssim t_{\rm mix}$.

Within this framework, the three classifications do not form a single required evolutionary sequence. Radial-gradient-only regions may trace weak or highly localized recent dilution. Regions identified by both methods may trace stronger recent dilution that produces a deficit relative to both metallicity expectations. The rMZR-only regions may instead trace gas that has mixed with the surrounding ISM while remaining under-enriched relative to the previous chemical evolution recorded by the underlying stellar population. The classifications may therefore reveal the strength and spatial extent of gas-flow events and how much the affected gas has mixed with the surrounding disk.

\subsection{Accreted Masses and Accretion Rate}\label{subsec:alm_mass_rate}
\begin{figure}[!h]
    \centering
    \includegraphics[width=\linewidth]{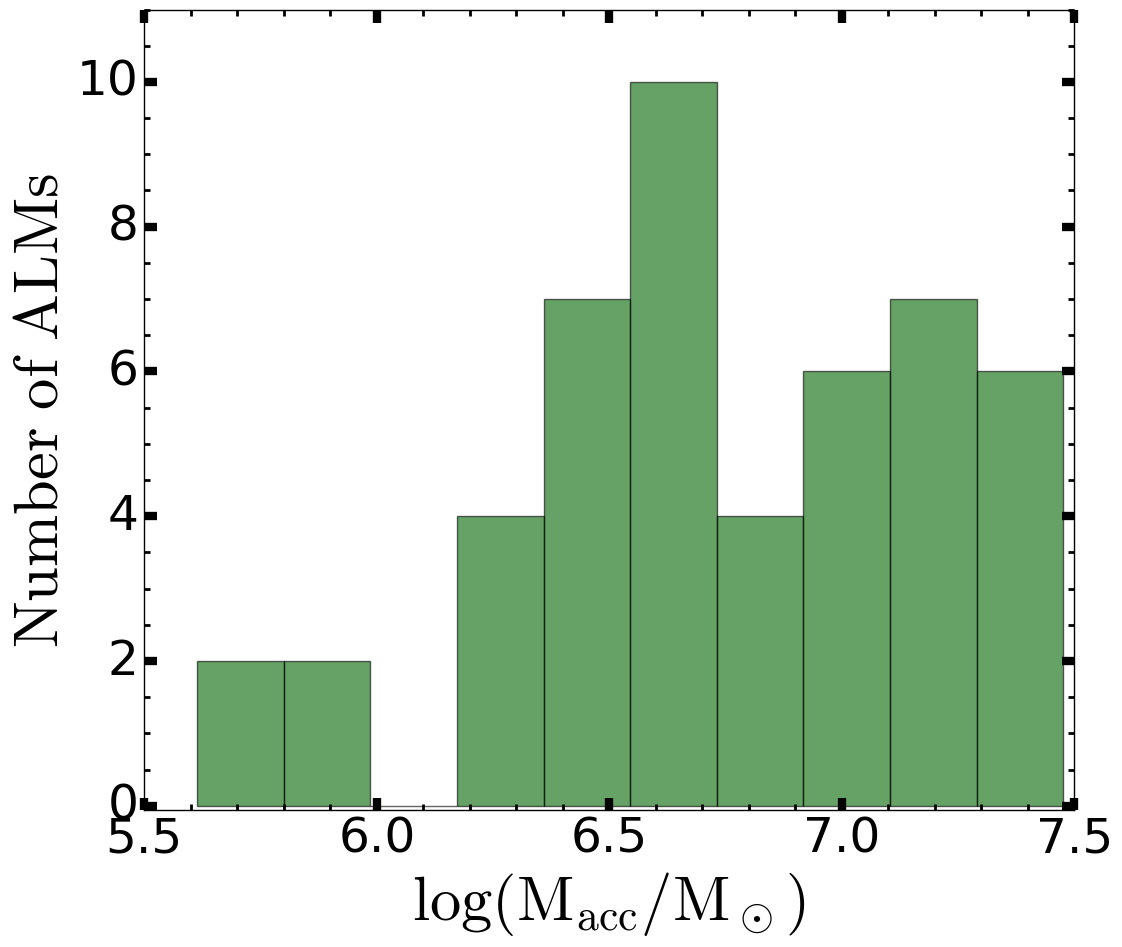}
    \caption{The distribution of estimated accreted masses required to dilute the 48 radial-gradient-selected ALM regions from their expected metallicities to their observed metallicities, assuming pristine accretion. The sample includes regions identified using the radial-gradient method alone or both the radial-gradient and rMZR methods, and excludes rMZR-only regions.}
    \label{fig:m_acc}
\end{figure}

We follow the procedure described in Section 4.3 of \citet{Olvera_2024ApJ...976..205O} to estimate the hydrogen mass that must be accreted to dilute an ALM region from its expected metallicity to its observed metallicity. For each region, we estimate the current hydrogen mass, $M_{\rm H,now}$, by integrating the VLA \ion{H}{1} surface-density map using the same masks discussed in Section~\ref{subsec:opt_img}. We denote the observed metallicity of the ALM region as $Z_{\rm obs}$ and estimate its original metallicity, $Z_{\rm exp}$, from the fitted radial metallicity gradient of its host galaxy at the region's galactocentric radius. We convert the oxygen abundance, expressed as $12+\log({\rm O/H})$, to an oxygen mass fraction using
\begin{equation}
    \log Z = \log({\rm O/H})+1.07.
\end{equation}
This calculation assumes that oxygen traces the metallicity of the gas and that the accreted gas mixes uniformly with the gas initially present in the region.

The accreted hydrogen mass is the difference between the current and original hydrogen masses,
\begin{equation}
    M_{\rm H,acc}=M_{\rm H,now}-M_{\rm H,original}.
\end{equation}
Assuming that the accreted gas is pristine, it contributes no oxygen, and the oxygen mass is conserved during dilution. Therefore,
\begin{equation}
    M_{\rm H,now}Z_{\rm obs}
    =
    M_{\rm H,original}Z_{\rm exp}.
\end{equation}
The original hydrogen mass and the mass required to produce the observed dilution are then
\begin{equation}
    M_{\rm H,original}
    =
    M_{\rm H,now}
    \left(\frac{Z_{\rm obs}}{Z_{\rm exp}}\right)
\end{equation}
and
\begin{equation}
    M_{\rm H,acc}
    =
    M_{\rm H,now}
    \left(1-\frac{Z_{\rm obs}}{Z_{\rm exp}}\right).
\end{equation}
Because gas with a nonzero metallicity would require more mass to produce the same amount of dilution, these estimates should be considered lower limits. For this calculation, we include the 29 ALM regions identified using both methods and the 19 regions identified using only the radial-gradient method. We exclude the rMZR-only regions because their metallicities are consistent with the measured radial gradients and do not necessarily indicate localized metallicity dilution.

Figure~\ref{fig:m_acc} shows the estimated accreted masses of the 48 ALM regions identified using the radial-gradient method, including those also identified by the rMZR. The masses span $5.6<\log(M_{\rm acc}/{\rm M_\odot})<7.5$, with the largest number of regions found near $10^{6.6}\ {\rm M_\odot}$ and a broader concentration between $10^{7.0}$ and $10^{7.5}\ {\rm M_\odot}$. This range is comparable to the masses of high-velocity clouds (HVCs) observed around the Milky Way \citep{Putman_2012ARA&A..50..491P} and those found in the TNG50 simulation \citep{Lucchini_2025ApJ...990..118L}. The comparable mass ranges are consistent with a scenario in which some ALM regions are produced by the accretion of HVC-like gas clouds, although the similarity in mass alone does not establish the origin or geometry of the inflow. We note that these regions would not be observationally confused with satellite galaxies given the difference in mass and size between the two populations.

To estimate an average accretion rate across the sample, we correct for the fraction of available \ion{H}{2} regions sampled by the spectroscopy. For each galaxy \textit{i}, we define
\begin{equation}
    C_i =
    \frac{N_{{\rm available},i}}
         {N_{{\rm fit},i}},
\end{equation}
where $N_{{\rm available},i}$ is the total number of candidate regions supplied to the BinoMask software and $N_{{\rm fit},i}$ is the number of regions used to measure the radial metallicity gradient, given in Table~3. We apply this correction separately to each galaxy because the fraction of available regions observed varies across the sample. The corrected accreted mass is then
\begin{equation}
    M_{{\rm acc,corr},i}
    =
    C_i
    \sum_{j\in{\rm radial\ selected}}M_{{\rm acc},ij},
\end{equation}
where the sum includes regions identified using either the radial-gradient method alone or both methods.

The 48 selected ALM regions have a combined estimated accreted mass of $4.12\times10^{8}\ {\rm M_\odot}$. After applying the galaxy-specific corrections, we obtain a total corrected mass of $2.80\times10^{9}\ {\rm M_\odot}$ across the sample. Dividing this value by the 11 galaxies gives an average corrected accreted mass of $2.55\times10^{8}\ {\rm M_\odot}$ per galaxy.

We assume that an ALM region remains identifiable for approximately one characteristic galactic dynamical time, which we take to be 250 Myr. The average hydrogen accretion rate is therefore
\begin{equation}
    \left\langle\dot{M}_{\rm acc}\right\rangle
    =
    \frac{1}{N_{\rm gal}}
    \sum_i
    \frac{M_{{\rm acc,corr},i}}{t_{\rm ALM}}
    =
    1.02
    \left(\frac{t_{\rm ALM}}{250\ {\rm Myr}}\right)^{-1}
    {\rm M_\odot\,yr^{-1}}.
\end{equation}

\begin{figure*}[t]
    \centering
    \includegraphics[width=\linewidth]{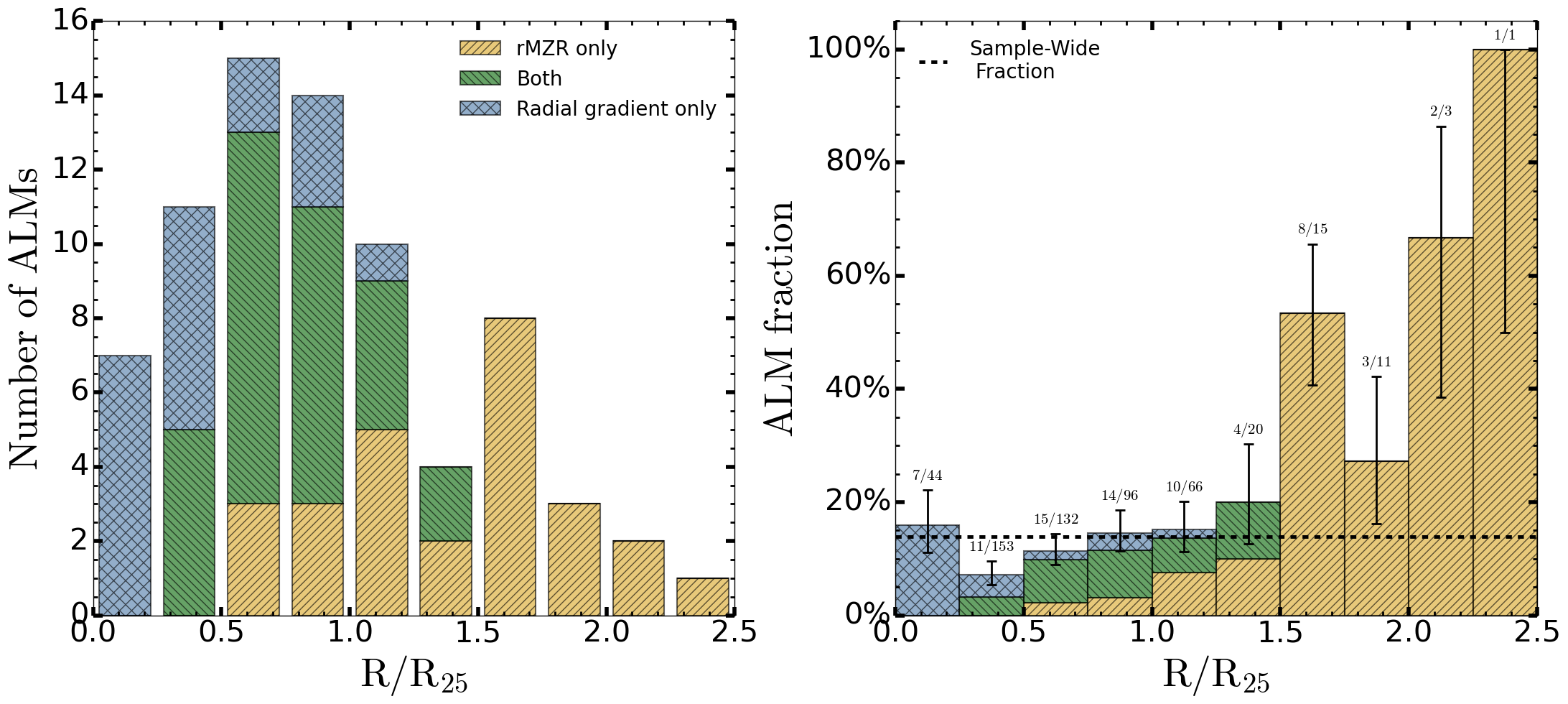}
    \caption{The galactocentric radial distribution of the 75 ALM regions, normalized by the optical radius, $R_{25}$, of each galaxy. The stacked bars separate regions identified using only the resolved mass--metallicity relation (rMZR; gold), using only the radial metallicity gradient (blue), and using both methods (green). The left panel shows the number of ALM regions in each radial bin, while the right panel shows the fraction of all observed \ion{H}{2} regions classified as ALMs. The numbers above the bars in the right panel give $N_{\rm ALM}/N_{\rm HII}$, and the error bars show the 68\% Wilson binomial confidence intervals. The dotted horizontal line marks the sample-wide ALM fraction of 13.9\%.}
    \label{fig:R_hist}
\end{figure*}

This estimate is comparable to the mean star formation rate of the galaxies in our sample, approximately $1.4\ {\rm M_\odot\,yr^{-1}}$. Under the adopted assumptions, the inferred supply of low-metallicity gas could therefore contribute substantially to replenishing the gas consumed by current star formation.

The number-based correction assumes that the observed regions are representative of the full catalog supplied to BinoMask. However, slit placement, target brightness, and mask geometry may affect which regions are observed. The inferred rate also depends on the assumed ALM lifetime and the accuracy of the gas masses assigned to individual regions. Since the \ion{H}{1} observations generally have lower spatial resolution than the optical spectra, nearby ALM regions may sample the same gas reservoir and may not represent independent accretion events. We therefore consider this value an approximate, time-averaged hydrogen accretion rate associated with the radial-gradient-selected ALM population rather than a direct measurement of the total inflow into these galaxies.

\subsection{Radial Position of ALM Regions}\label{subsec:alm_radius}

The left panel of Figure \ref{fig:R_hist} shows the radial distribution of the 75 ALM regions, separated by the method used to identify them, while the right panel shows the fraction of the 541 \ion{H}{2} regions retained for this analysis that are classified as ALMs in each radial bin. These panels provide complementary views of the radial distribution. The number of ALM regions indicates where most of the detected population is located, whereas the ALM fraction measures how frequently ALMs occur relative to the number of \ion{H}{2} regions observed at each radius.

The ALM regions span $0.11<\rm R/R_{25}<2.37$, with a median radius of $\rm R/R_{25}=0.84$. Of the 75 ALM regions, 47 are found within $\rm R_{25}$ and 28 are found beyond $\rm R_{25}$. Therefore, most of the detected ALM regions are located within the optical disk, although approximately 37\% are found in the extended disk. The number distribution peaks between approximately $0.5<\rm R/R_{25}<1.0$ and then decreases toward larger radii.

The fractional distribution shows a different, but complementary, radial trend. Within $\rm R_{25}$, 47 of the 425 observed \ion{H}{2} regions are classified as ALMs, corresponding to an ALM fraction of 11.1\%. Beyond $\rm R_{25}$, 28 of 116 regions are classified as ALMs, giving an ALM fraction of 24.1\%. Thus, although the total number of ALM regions is larger within the optical disk, an observed outer-disk \ion{H}{2} region is approximately 2.2 times more likely to be classified as an ALM. The increase is especially apparent beyond $1.5\rm R_{25}$, where 14 of the 30 observed regions are classified as ALMs. However, the number of observed \ion{H}{2} regions becomes small at these radii, resulting in large binomial uncertainties. In particular, the fractions in the two outermost bins are based on only three and one observed regions, respectively.

The classification breakdown provides important context for interpreting this increase in the outer-disk ALM fraction. Of the 75 ALM regions, 27 are identified using only the rMZR, 19 are identified using only the radial metallicity gradient, and 29 are identified using both methods. The radial-gradient-only regions are mainly found in the inner disk, while regions identified using both methods are mostly found at intermediate radii. In comparison, the outer-disk population is dominated by rMZR-only regions. Of the 28 ALM regions beyond $R_{25}$, 21 are identified using only the rMZR, six are identified using both methods, and one is identified using only the radial gradient. All 14 ALM regions beyond $1.5\rm R_{25}$ are identified only by the rMZR. The increase in the total ALM fraction at the largest radii is therefore driven entirely by the rMZR-only population.

As discussed in Section~\ref{subsec:method_comparison}, this difference is related to the expected metallicity used by each method. The rMZR does not explicitly account for the radial metallicity gradient of each galaxy. Since metallicity and stellar mass surface density generally decrease with radius, some outer-disk regions may fall below the sample-wide rMZR while remaining consistent with the radial abundance profile of their host galaxy. The rMZR-only population at large radii may therefore include both localized metallicity anomalies and regions produced by differences between the rMZR and the metallicity structure of individual disks. In comparison, regions identified using both methods are metal poor relative to both their galactocentric positions and local stellar mass surface densities. These regions, which are mostly found within or near $R_{25}$, provide the strongest evidence for localized metallicity dilution.

Previous studies have also found that the incidence of ALM gas depends on galactocentric radius. Using MaNGA observations, \citet{Hwang_2019ApJ...872..144H} found that the fraction of regions containing ALM gas increases toward larger radii. Our fractional distribution is qualitatively consistent with this result because the ALM fraction is higher beyond $R_{25}$ than within the optical disk. At the same time, the number distribution shows that most of the ALM regions in our sample are located within or near $R_{25}$. These results are not contradictory: the inner disk contains more observed \ion{H}{2} regions and therefore more ALMs in total, while ALMs make up a larger fraction of the more sparsely sampled outer-disk population. Differences in sample selection, spatial resolution, radial coverage, and the methods used to identify ALM gas may affect the strength of the radial trend.

The number and fractional distributions may reflect different manifestations of low-metallicity gas within galaxy disks. The ALM regions found within $R_{25}$ may be produced by metal-poor gas reaching the inner disk through vertical accretion, such as the infall of HVC-like clouds, or through the cooling and mixing of low-metallicity circumgalactic gas with recycled material in a galactic fountain \citep{Putman_2012ARA&A..50..491P, Werk_2011ApJ...735...71W, Bresolin_2012ApJ...750..122B}. Numerical simulations show that recently accreted metal-poor gas can form off-center, H$\alpha$-bright regions with metallicities below that of the surrounding interstellar medium \citep{Ceverino_2016MNRAS.457.2605C}. The location at which this gas enters the disk depends on the geometry and angular momentum of the inflow. After entering the disk, the gas may also be redistributed by spiral arms, bars, or turbulence. The elevated outer-disk ALM fraction may similarly reflect the greater susceptibility of the chemically less evolved outer disk to low-metallicity gas accretion. However, because this increase is driven by rMZR-only regions, it is difficult to determine whether it represents localized metallicity dilution or differences between the sample-wide rMZR and the underlying metallicity structure of individual galaxies.

Overall, most of the detected ALM regions are located within or near $R_{25}$, but ALMs constitute a larger fraction of the observed \ion{H}{2} regions in the outer disk. The strongest localized metallicity anomalies, identified using both methods, are concentrated within or near the optical disk, whereas the elevated outer fraction is driven by rMZR-only regions. The number and fractional distributions therefore provide complementary evidence that the incidence and nature of low-metallicity gas vary across galaxy disks.

\subsection{Connection Between Disk Metallicity Structure and Circumgalactic Ly$\alpha$ Absorption}
\label{sec:disk_cgm}

Previous studies have shown that circumgalactic Ly$\alpha$ absorption is connected to the gas content and star-forming properties of galaxies. \citet{borthakur2016} found that impact-parameter-corrected Ly$\alpha$ equivalent width correlates with specific star formation rate, while \citet{borthakur2024diisc} found that Ly$\alpha$ equivalent width is more closely related to distance from the \ion{H}{1} disk than to distance normalized by the virial radius. The latter study also found positive correlations between Ly$\alpha$ excess and both \ion{H}{1} mass and gas fraction, indicating that circumgalactic Ly$\alpha$ absorption is related to the global neutral-gas reservoir. The QSO sightlines observed with the Cosmic Origins Spectrograph (COS) aboard the Hubble Space Telescope (HST) provide a complementary probe of cool circumgalactic gas that is not generally available in samples selected for spatially resolved disk spectroscopy. These observations allow us to test whether galaxies with larger Ly$\alpha$ excesses also have flatter metallicity gradients or a higher incidence of ALM regions, as might be expected if the detected circumgalactic gas is connected to recent accretion and chemical dilution in the disk. Conversely, the absence of a correlation may indicate that this connection depends on the spatial distribution, geometry, or evolutionary timescale of the accreting gas.

We adopt the observed Ly$\alpha$ equivalent widths and associated limits from \citet{borthakur2024diisc}. Because Ly$\alpha$ equivalent width depends strongly on the projected distance of the QSO sightline from the galaxy, we calculate the Ly$\alpha$ excess after removing the expected radial dependence using relations based on both the virial radius and \ion{H}{1} radius. The first relation is based on the virial-radius-normalized impact parameter from \citet{borthakur2016},
\begin{equation}
    \log W_{\mathrm{Ly}\alpha,\mathrm{expected}}
    =
    -0.387\left(\frac{\rho}{R_{\mathrm{vir}}}\right)-0.044,
\end{equation}
while the second uses the \ion{H}{1}-radius-normalized relation from \citet{borthakur2024diisc},
\begin{equation}
    \log W_{\mathrm{Ly}\alpha,\mathrm{expected}}
    =
    -0.6810\log\left(\frac{\rho}{R_{\mathrm{HI}}}\right)+0.1894.
\end{equation}
For each relation, we define the Ly$\alpha$ excess as
\begin{equation}
    \Delta\log W_{\mathrm{Ly}\alpha}
    =
    \log W_{\mathrm{Ly}\alpha,\mathrm{observed}}
    -
    \log W_{\mathrm{Ly}\alpha,\mathrm{expected}}.
\end{equation}
A positive value represents stronger Ly$\alpha$ absorption than expected at the normalized impact parameter of the sightline, while a negative value represents weaker absorption.

NGC~5669 is excluded from the Ly$\alpha$ analysis because no equivalent-width measurement is available. Of the remaining ten galaxies, eight have Ly$\alpha$ detections, while NGC~4321 has an upper limit and NGC~3344 has a lower limit. Considering only the eight detections, the $R_{\mathrm{HI}}$-normalized excess has a nominal positive slope with metallicity gradient of $0.67\pm0.61$, but the relationship is not statistically significant ($p\simeq0.32$). The $R_{\mathrm{vir}}$-normalized excess is effectively independent of metallicity gradient, with a slope of $0.009\pm0.754$ and $p=0.99$. We also account for the two limiting measurements using an interval-censored Gaussian regression. The resulting slopes are $-0.96\pm1.00$ for the $R_{\mathrm{HI}}$-normalized excess and $-1.50\pm1.02$ for the $R_{\mathrm{vir}}$-normalized excess, with $p=0.34$ and $p=0.14$, respectively. The change in sign relative to the detection-only fits is primarily driven by the locations of the two censored measurements: NGC~4321 has a relatively flat, slightly positive metallicity gradient and a low upper limit on Ly$\alpha$ excess, while NGC~3344 has the steepest negative gradient in the sample and a lower limit on its excess. In both cases, the uncertainties encompass negative, zero, and positive slopes, indicating that the direction of any relationship is not robustly constrained.

We next compare the incidence of ALM regions with the metallicity gradient and Ly$\alpha$ excess. For this analysis, $f_{\mathrm{ALM}}$ is defined as the number of regions identified by the radial-gradient method, either independently or by both the radial-gradient and rMZR methods, divided by the total number of observed star-forming regions in each galaxy. We find no significant correlation between $f_{\mathrm{ALM}}$ and metallicity gradient, $R_{\mathrm{HI}}$-normalized Ly$\alpha$ excess, or $R_{\mathrm{vir}}$-normalized Ly$\alpha$ excess. The Pearson correlation coefficients are $r=-0.02$, $r\simeq0.00$, and $r=-0.01$, respectively, while the corresponding Spearman coefficients are $\rho=0.26$, $\rho=0.21$, and $\rho=0.31$. All three relationships have $p>0.4$. Because the limiting measurements occur in the independent variable, the Ly$\alpha$ excess correlations are calculated using the eight galaxies with detected equivalent widths.

The absence of a significant relationship does not necessarily imply that ALM regions and circumgalactic gas have unrelated origins. Ly$\alpha$ absorption is measured along a single QSO sightline per galaxy and may be sensitive to the local density, covering fraction, orientation, and velocity structure of the neutral gas, whereas the metallicity gradient and $f_{\mathrm{ALM}}$ combine measurements distributed across the disk. Gas associated with an individual ALM region may occupy only a small portion of the CGM and may not intersect the available sightline. The metallicity signature of accretion also depends on the mass and metallicity of the incoming gas, the properties of the pre-existing ISM, and the timescales for spatial mixing and chemical enrichment. We therefore find no significant evidence that the global metallicity gradient or $f_{\mathrm{ALM}}$ is related to Ly$\alpha$ excess in the present sample, although the small sample size, censored measurements, and availability of only one CGM sightline per galaxy limit the strength of this conclusion. A larger sample with multiple sightlines around individual galaxies will be needed to test whether localized metallicity dilution is associated with enhanced neutral-gas absorption at particular locations along the disk--CGM interface.

\section{Conclusion} \label{sec:conclusion}

In this work, we investigate the properties of 542 H$\alpha$-bright regions across 11 nearby star-forming galaxies in the DIISC survey. We use optical spectroscopy from Binospec, optical imaging from the VATT, and \ion{H}{1} 21 cm imaging from the VLA to search for indirect evidence of recent low-metallicity gas accretion. We measure the gas-phase metallicity of each region using the N2 metallicity indicator and identify ALM regions by comparing their measured metallicities with those expected from both their galactocentric positions and local stellar mass surface densities. We then investigate the possible physical differences between these ALM classifications, estimate the mass required to produce the observed metallicity dilution, examine where the ALM regions are found within their host galaxies, and test for a connection between chemical structure within the disk and gas in the CGM. Our main results are summarized below:

\begin{enumerate}

    \item \textbf{The DIISC galaxies have typical metallicity gradients.} Most galaxies have negative gradients that are consistent with those of nearby star-forming galaxies. The strongest metallicity outliers remain anomalous when alternative strong-line metallicity indicators are used.

    \item \textbf{We identify 75 ALM regions.} The radial-gradient and rMZR methods identify 48 and 56 regions, respectively. The combined sample contains 19 radial-gradient-only regions, 27 rMZR-only regions, and 29 regions identified using both methods.
    
    \item \textbf{Both methods identify the strongest candidates.} Regions identified using both methods are metal poor relative to their galactocentric positions and local stellar mass surface densities, providing the strongest evidence for localized metallicity dilution.
    
    \item \textbf{The classifications may trace strength, spatial extent, and degree of mixing of chemically distinct gas.} Radial-gradient-only regions may represent recent, localized metallicity dilution, while rMZR-only regions may trace gas that has mixed with the surrounding ISM but remains under-enriched relative to the chemical history recorded by the local stellar population.
    
    \item \textbf{ALM masses are comparable to HVC masses.} The estimated accreted masses span $5.6<\log(M_{\rm acc}/{\rm M_\odot})<7.5$, consistent with the mass range of observed and simulated HVCs. 
    
    \item \textbf{The inferred accretion rate is comparable to the SFR.} We estimate a mean hydrogen accretion rate of $1.02\ {\rm M_\odot\,yr^{-1}}$ per galaxy for an assumed ALM lifetime of 250 Myr using the ALM identified by the radial gradient method. This is comparable to the sample's mean SFR of approximately $1.4\ {\rm M_\odot\,yr^{-1}}$.
    
    \item \textbf{ALMs occur throughout disks but are more common at large radii.} Although 47 of the 75 ALMs are found within $R_{25}$, the ALM fraction increases from 11.1\% within $R_{25}$ to 24.1\% beyond it. This elevated outer-disk fraction is dominated by rMZR-only regions, while regions identified using both methods are mostly found within or near $R_{25}$.

    \item \textbf{No evidence for connection between CGM and ALM regions.} We find no significant correlation between Ly\(\alpha\) excess and either the metallicity gradient or the fraction of radial-gradient-selected ALM regions, although the comparison is limited by the small sample and the availability of only one CGM sightline per galaxy. 

\end{enumerate}

Together, these results show that ALM regions are not a single uniform population. Their metallicities, radial positions, and classifications may reflect differences in the strength and spatial extent of dilution and the degree to which chemically distinct gas has mixed with the surrounding ISM. Regions identified by both the radial-gradient and rMZR methods may provide the strongest evidence for recent, substantial, and localized dilution. Radial-gradient-only regions may instead trace weaker or more localized metallicity variations, while rMZR-only regions may trace gas that has mixed with the surrounding ISM but remains under-enriched relative to the previous chemical evolution recorded by the local stellar population. These interpretations provide a possible evolutionary framework, but the classifications alone do not establish the age of an accretion event or demonstrate that the populations form a temporal sequence.

Additional observations and analysis are needed to test this proposed evolution. Integral field unit observations would constrain the full spatial extent and morphology of the ALM regions, reveal whether nearby detections are components of the same chemically distinct structure, and measure their metallicity contrast with the immediately surrounding ISM. Spatially resolved ionized- and neutral-gas kinematics could test whether rMZR-only regions are associated with enhanced turbulence or more efficient mixing, while stellar-population measurements could determine whether their underlying stars indicate a greater enrichment history than the present gas. Combining these measurements with local orbital periods and comparisons to simulations would help determine whether radial-gradient-only, both-method, and rMZR-only regions represent different stages of dilution and mixing. Similar studies at higher redshift will be needed to determine how the frequency, spatial distribution, mass, and physical origin of ALM regions evolve across cosmic time.

\begin{acknowledgments}
We thank the extended DIISC collaboration for their contributions to the program.

This material is based upon work supported by the National Science Foundation Graduate Research Fellowship Program under Grant No. 2233001. Any opinions, findings, and conclusions or recommendations expressed in this material are those of the author(s) and do
not necessarily reflect the views of the National Science Foundation.

A.O. and S.B. acknowledge support from the NSF grant 2511242.

The Arizona State University authors acknowledge the twenty-three Native Nations that have inhabited this land for centuries. Arizona State University's four campuses are located in the Salt River Valley on ancestral territories of Indigenous peoples, including the Akimel O’odham (Pima) and Pee Posh (Maricopa) Indian Communities, whose care and keeping of these lands allow us to be here today.

We thank the support staff at the MMT Observatory, the Steward Observatory, the Vatican Advanced Technology Telescope, the Very Large Array, the National Radio Astronomy Observatory, and the Space Telescope Science Institute for their help with this project.

Observations reported here were obtained at the MMT Observatory, a joint facility of the Smithsonian Institution and the University of Arizona.

This work is also partly based on observations with the VATT: the Alice P. Lennon Telescope and the Thomas J. Bannan Astrophysics Facility.

The National Radio Astronomy Observatory is a facility of the National Science Foundation operated under cooperative agreement by Associated Universities, Inc.
\end{acknowledgments}





%
\facilities{HST, MMT (Binospec), VATT, and VLA}

\software{astropy \citep{astropy:2013,astropy:2018,astropy:2022},  
          \texttt{ipython/ jupyter} \citep{Perez_2007CSE.....9c..21P,Kluyver_2016ppap.book...87K}, 
          \texttt{matplotlib} \citep{Hunter:2007},
          \texttt{NumPy} \citep{harris2020array},
          photutils \citep{Bradley_larry_bradley_2023_7946442},
          \texttt{pPXF} \citep{Cappellari_2023MNRAS.526.3273C},
          \texttt{specutils} \citep{specutils}, and 
          \texttt{Python} from \url{https://www.python.org}
          }


\appendix

\section{Alternative Metallicity Indicators}\label{appx:AltIndicators}

\begin{figure}[!h]
    \centering
    \includegraphics[width=\linewidth]{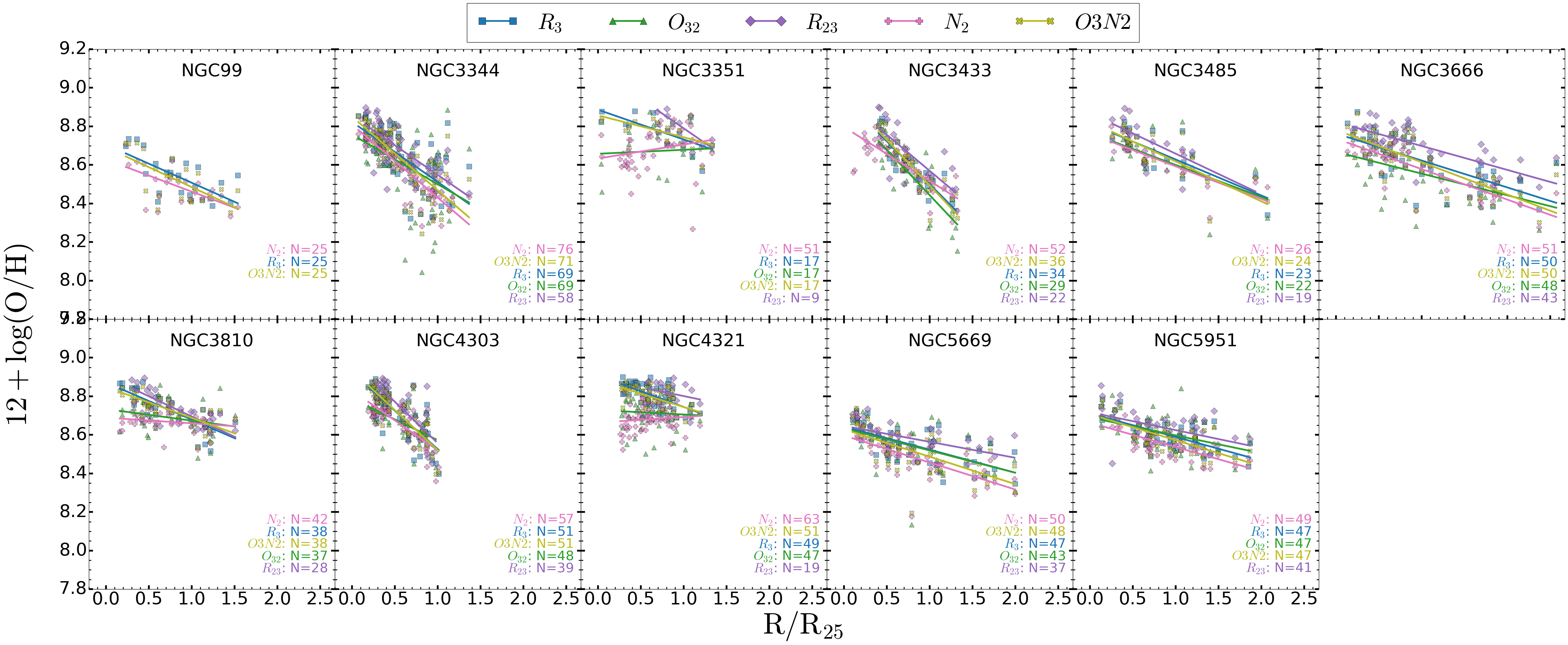}
    \caption{Each panel shows how the gas-phase metallicity in each galaxy varies as a function of galactocentric radius. The different solid line colors and point shapes represent different metallicity indicators. The text in the bottom right of each panel gives the number of regions used in the line fit of each region.}
    \label{fig:alt_metallicity}
\end{figure}

Different strong-line metallicity indicators have produced differing metallicities and radial gradients \citep{Kewley_2008ApJ...681.1183K, Poetrodjojo_2019MNRAS.487...79P}. In addition, \cite{Kreckel_2019ApJ...887...80K} discussed how different metallicity indicators result in a differing number of outlier regions. They reported that N2 and O3N2 showed a 20\% and 10\%, respectively, outlier overlap with the Scal indicator \citep{Pilyugin_2016MNRAS.457.3678P}. In order to test whether our choice of metallicity indicator affects ALM identification, we measure each galaxy's radial metallicity gradient using five other indicators: $\rm R_2 = log($[\ion{O}{2}]$\lambda~3727/{\rm H}\beta)$, $\rm R_3 = log($[\ion{O}{3}]$\lambda~5007/{\rm H}\beta)$, $\rm O_{32} = log($[\ion{O}{3}]$\lambda~5007/$[\ion{O}{2}]$\lambda~3727)$, $\rm R_{23} = log($\{[\ion{O}{2}]$\lambda~3727$+[\ion{O}{3}]$\lambda~5007\}/{\rm H}\beta)$, and $\rm O3N2 = log($\{[\ion{O}{3}]$\lambda~5007/{\rm H}\beta\}$/\{[\ion{N}{2}]$\lambda~6583$/H$\alpha$\}). The additional emission lines mentioned are measured using the same procedure as mentioned in Section \ref{subsec:opt_spec}. 

Figure \ref{fig:alt_metallicity} shows each galaxy's radial metallicity gradient from the different metallicity indicators and reveals that the gradient in each galaxy remains broadly consistent. On the other hand, the metallicity indicators have differing intercept values and absolute metallicities per region. The $\rm R_{23}$ indicator tends to estimate the highest metallicities while the N2 indicator yields the lowest metallicities. The other indicators predict metallicity values between $\rm R_{23}$ and N2. While absolute metallicity values tend to shift to lower or higher values depending on which indicators are used, the metallicity deviation sign and magnitude of each region remains roughly the same for large fraction of the regions in each galaxy.

NGC~3351 is the biggest except from the previous statements. The [\ion{O}{3}] line is not present in a majority of the spectra from NGC~3351, so the five [\ion{O}{3}]-based indicators suffer from low number statistics making it challenging to properly compare different indicators.

Overall, we find that the metallicity indicator does not affect our ALM measurement to a significant degree. The strongest ALM regions remain anomalous in metallicity in multiple indicators. Therefore, we use the N2 indicator as our primary metallicity proxy throughout the paper.


\bibliography{sample701}{}

@ARTICLE{CASA_2022PASP..134k4501C,
       author = {{CASA Team} and {Bean}, Ben and {Bhatnagar}, Sanjay and {Castro}, Sandra and {Donovan Meyer}, Jennifer and {Emonts}, Bjorn and {Garcia}, Enrique and {Garwood}, Robert and {Golap}, Kumar and {Gonzalez Villalba}, Justo and {Harris}, Pamela and {Hayashi}, Yohei and {Hoskins}, Josh and {Hsieh}, Mingyu and {Jagannathan}, Preshanth and {Kawasaki}, Wataru and {Keimpema}, Aard and {Kettenis}, Mark and {Lopez}, Jorge and {Marvil}, Joshua and {Masters}, Joseph and {McNichols}, Andrew and {Mehringer}, David and {Miel}, Renaud and {Moellenbrock}, George and {Montesino}, Federico and {Nakazato}, Takeshi and {Ott}, Juergen and {Petry}, Dirk and {Pokorny}, Martin and {Raba}, Ryan and {Rau}, Urvashi and {Schiebel}, Darrell and {Schweighart}, Neal and {Sekhar}, Srikrishna and {Shimada}, Kazuhiko and {Small}, Des and {Steeb}, Jan-Willem and {Sugimoto}, Kanako and {Suoranta}, Ville and {Tsutsumi}, Takahiro and {van Bemmel}, Ilse M. and {Verkouter}, Marjolein and {Wells}, Akeem and {Xiong}, Wei and {Szomoru}, Arpad and {Griffith}, Morgan and {Glendenning}, Brian and {Kern}, Jeff},
        title = "{CASA, the Common Astronomy Software Applications for Radio Astronomy}",
      journal = {\pasp},
         year = 2022,
        month = nov,
       volume = {134},
       number = {1041},
          eid = {114501},
        pages = {114501},
          doi = {10.1088/1538-3873/ac9642},
archivePrefix = {arXiv},
       eprint = {2210.02276},
 primaryClass = {astro-ph.IM},
       adsurl = {https://ui.adsabs.harvard.edu/abs/2022PASP..134k4501C}
}

@ARTICLE{Padave_2024ApJ...960...24P,
       author = {{Padave}, Mansi and {Borthakur}, Sanchayeeta and {Gim}, Hansung B. and {Thilker}, David and {Jansen}, Rolf A. and {Monkiewicz}, Jacqueline and {Kennicutt}, Robert C. and {Kauffmann}, Guinevere and {Fox}, Andrew J. and {Momjian}, Emmanuel and {Heckman}, Timothy},
        title = "{DIISC-III. Signatures of Stellar Disk Growth in Nearby Galaxies}",
      journal = {\apj},
         year = 2024,
        month = jan,
       volume = {960},
       number = {1},
          eid = {24},
        pages = {24},
          doi = {10.3847/1538-4357/ad029b},
archivePrefix = {arXiv},
       eprint = {2310.08482},
 primaryClass = {astro-ph.GA},
       adsurl = {https://ui.adsabs.harvard.edu/abs/2024ApJ...960...24P}
}

@ARTICLE{Bertin_1996A&AS..117..393B,
       author = {{Bertin}, E. and {Arnouts}, S.},
        title = "{SExtractor: Software for source extraction.}",
      journal = {\aaps},
         year = "1996",
        month = "Jun",
       volume = {117},
        pages = {393-404},
          doi = {10.1051/aas:1996164},
       adsurl = {https://ui.adsabs.harvard.edu/abs/1996A&AS..117..393B}
}

@ARTICLE{Padave_2025ApJ...986..145P,
       author = {{Padave}, Mansi and {Borthakur}, Sanchayeeta and {Jansen}, Rolf A. and {Thilker}, David A. and {Monkiewicz}, Jacqueline and {Windhorst}, Rogier A.},
        title = "{DIISC-V: Variations in H{\ensuremath{\alpha}}-to-FUV Star Formation Rate Ratios Across Star-forming Regions in Nearby Galaxies}",
      journal = {\apj},
         year = 2025,
        month = jun,
       volume = {986},
       number = {2},
          eid = {145},
        pages = {145},
          doi = {10.3847/1538-4357/adce7b},
archivePrefix = {arXiv},
       eprint = {2407.16690},
 primaryClass = {astro-ph.GA},
       adsurl = {https://ui.adsabs.harvard.edu/abs/2025ApJ...986..145P}
}

@ARTICLE{kansky_2019PASP..131g5005K,
       author = {{Kansky}, Jan and {Chilingarian}, Igor and {Fabricant}, Daniel and {Matthews}, Anne and {Moran}, Sean and {Paegert}, Martin and {Duane Gibson}, J. and {Porter}, Dallan and {Roll}, John},
        title = "{Binospec Software System}",
      journal = {\pasp},
         year = 2019,
        month = jul,
       volume = {131},
       number = {1001},
        pages = {075005},
          doi = {10.1088/1538-3873/ab1ceb},
archivePrefix = {arXiv},
       eprint = {1905.03321},
 primaryClass = {astro-ph.IM},
       adsurl = {https://ui.adsabs.harvard.edu/abs/2019PASP..131g5005K}
}

@ARTICLE{Olvera_2024ApJ...976..205O,
       author = {{Olvera}, Alejandro J. and {Borthakur}, Sanchayeeta and {Padave}, Mansi and {Heckman}, Timothy and {Gim}, Hansung B. and {Koplitz}, Brad and {Dupuis}, Christopher and {Momjian}, Emmanuel and {Jansen}, Rolf A.},
        title = "{DIISC-IV. DIISCovery of Anomalously Low Metallicity H II Regions in NGC 99: Indirect Evidence of Gas Inflows}",
      journal = {\apj},
         year = 2024,
        month = dec,
       volume = {976},
       number = {2},
          eid = {205},
        pages = {205},
          doi = {10.3847/1538-4357/ad8238},
archivePrefix = {arXiv},
       eprint = {2408.08303},
 primaryClass = {astro-ph.GA},
       adsurl = {https://ui.adsabs.harvard.edu/abs/2024ApJ...976..205O}
}

@ARTICLE{Cappellari_2023MNRAS.526.3273C,
       author = {{Cappellari}, Michele},
        title = "{Full spectrum fitting with photometry in PPXF: stellar population versus dynamical masses, non-parametric star formation history and metallicity for 3200 LEGA-C galaxies at redshift z {\ensuremath{\approx}} 0.8}",
      journal = {\mnras},
         year = 2023,
        month = dec,
       volume = {526},
       number = {3},
        pages = {3273-3300},
          doi = {10.1093/mnras/stad2597},
archivePrefix = {arXiv},
       eprint = {2208.14974},
 primaryClass = {astro-ph.GA},
       adsurl = {https://ui.adsabs.harvard.edu/abs/2023MNRAS.526.3273C}
}

@ARTICLE{Vazdekis_2016MNRAS.463.3409V,
       author = {{Vazdekis}, A. and {Koleva}, M. and {Ricciardelli}, E. and {R{\"o}ck}, B. and {Falc{\'o}n-Barroso}, J.},
        title = "{UV-extended E-MILES stellar population models: young components in massive early-type galaxies}",
      journal = {\mnras},
         year = 2016,
        month = dec,
       volume = {463},
       number = {4},
        pages = {3409-3436},
          doi = {10.1093/mnras/stw2231},
archivePrefix = {arXiv},
       eprint = {1612.01187},
 primaryClass = {astro-ph.GA},
       adsurl = {https://ui.adsabs.harvard.edu/abs/2016MNRAS.463.3409V}
}

@ARTICLE{Salpeter_1955ApJ...121..161S,
       author = {{Salpeter}, Edwin E.},
        title = "{The Luminosity Function and Stellar Evolution}",
      journal = {ApJ},
         year = 1955,
       volume = {121},
        pages = {161},
          doi = {10.1086/145971}
}

@ARTICLE{Curti_FMR_2020MNRAS.491..944C,
       author = {{Curti}, Mirko and {Mannucci}, Filippo and {Cresci}, Giovanni and {Maiolino}, Roberto},
        title = "{The mass-metallicity and the fundamental metallicity relation revisited on a fully T$_{e}$-based abundance scale for galaxies}",
      journal = {\mnras},
         year = 2020,
        month = jan,
       volume = {491},
       number = {1},
        pages = {944-964},
          doi = {10.1093/mnras/stz2910},
archivePrefix = {arXiv},
       eprint = {1910.00597},
 primaryClass = {astro-ph.GA},
       adsurl = {https://ui.adsabs.harvard.edu/abs/2020MNRAS.491..944C}
}

@software{Newville_2021zndo....598352N,
       author = {{Newville}, Matthew and {Otten}, Renee and {Nelson}, Andrew and {Stensitzki}, Till and {Ingargiola}, Antonino and {Allan}, Daniel and {Fox}, Austin and {Carter}, Faustin and {Rawlik}, Michal},
        title = "{LMFIT: Non-Linear Least-Squares Minimization and Curve-Fitting for Python}",
         year = 2025,
        month = mar,
          eid = {10.5281/zenodo.598352},
          doi = {10.5281/zenodo.598352},
      version = {1.3.3},
    publisher = {Zenodo},
       adsurl = {https://ui.adsabs.harvard.edu/abs/2021zndo....598352N}
}

@ARTICLE{Zaritsky_1994ApJ...420...87Z,
       author = {{Zaritsky}, Dennis and {Kennicutt}, Jr., Robert C. and {Huchra}, John P.},
        title = "{H II Regions and the Abundance Properties of Spiral Galaxies}",
      journal = {\apj},
         year = 1994,
        month = jan,
       volume = {420},
        pages = {87},
          doi = {10.1086/173544},
       adsurl = {https://ui.adsabs.harvard.edu/abs/1994ApJ...420...87Z}
}

@ARTICLE{Sanchez_2014A&A...563A..49S,
       author = {{S{\'a}nchez}, S.~F. and {Rosales-Ortega}, F.~F. and {Iglesias-P{\'a}ramo}, J. and {Moll{\'a}}, M. and {Barrera-Ballesteros}, J. and {Marino}, R.~A. and {P{\'e}rez}, E. and {S{\'a}nchez-Blazquez}, P. and {Gonz{\'a}lez Delgado}, R. and {Cid Fernandes}, R. and {de Lorenzo-C{\'a}ceres}, A. and {Mendez-Abreu}, J. and {Galbany}, L. and {Falcon-Barroso}, J. and {Miralles-Caballero}, D. and {Husemann}, B. and {Garc{\'\i}a-Benito}, R. and {Mast}, D. and {Walcher}, C.~J. and {Gil de Paz}, A. and {Garc{\'\i}a-Lorenzo}, B. and {Jungwiert}, B. and {V{\'\i}lchez}, J.~M. and {J{\'\i}lkov{\'a}}, Lucie and {Lyubenova}, M. and {Cortijo-Ferrero}, C. and {D{\'\i}az}, A.~I. and {Wisotzki}, L. and {M{\'a}rquez}, I. and {Bland-Hawthorn}, J. and {Ellis}, S. and {van de Ven}, G. and {Jahnke}, K. and {Papaderos}, P. and {Gomes}, J.~M. and {Mendoza}, M.~A. and {L{\'o}pez-S{\'a}nchez}, {\'A}. R.},
        title = "{A characteristic oxygen abundance gradient in galaxy disks unveiled with CALIFA}",
      journal = {\aap},
         year = 2014,
        month = mar,
       volume = {563},
          eid = {A49},
        pages = {A49},
          doi = {10.1051/0004-6361/201322343},
archivePrefix = {arXiv},
       eprint = {1311.7052},
 primaryClass = {astro-ph.CO},
       adsurl = {https://ui.adsabs.harvard.edu/abs/2014A&A...563A..49S}
}

@ARTICLE{Ho_2015MNRAS.448.2030H,
       author = {{Ho}, I.-Ting and {Kudritzki}, Rolf-Peter and {Kewley}, Lisa J. and {Zahid}, H. Jabran and {Dopita}, Michael A. and {Bresolin}, Fabio and {Rupke}, David S.~N.},
        title = "{Metallicity gradients in local field star-forming galaxies: insights on inflows, outflows, and the coevolution of gas, stars and metals}",
      journal = {\mnras},
         year = 2015,
        month = apr,
       volume = {448},
       number = {3},
        pages = {2030-2054},
          doi = {10.1093/mnras/stv067},
archivePrefix = {arXiv},
       eprint = {1501.02668},
 primaryClass = {astro-ph.GA},
       adsurl = {https://ui.adsabs.harvard.edu/abs/2015MNRAS.448.2030H}
}

@ARTICLE{Sanchez-Menguiano_2016A&A...587A..70S,
       author = {{S{\'a}nchez-Menguiano}, L. and {S{\'a}nchez}, S.~F. and {P{\'e}rez}, I. and {Garc{\'\i}a-Benito}, R. and {Husemann}, B. and {Mast}, D. and {Mendoza}, A. and {Ruiz-Lara}, T. and {Ascasibar}, Y. and {Bland-Hawthorn}, J. and {Cavichia}, O. and {D{\'\i}az}, A.~I. and {Florido}, E. and {Galbany}, L. and {G{\'o}nzalez Delgado}, R.~M. and {Kehrig}, C. and {Marino}, R.~A. and {M{\'a}rquez}, I. and {Masegosa}, J. and {M{\'e}ndez-Abreu}, J. and {Moll{\'a}}, M. and {Del Olmo}, A. and {P{\'e}rez}, E. and {S{\'a}nchez-Bl{\'a}zquez}, P. and {Stanishev}, V. and {Walcher}, C.~J. and {L{\'o}pez-S{\'a}nchez}, {\'A}. R. and {CALIFA Collaboration}},
        title = "{Shape of the oxygen abundance profiles in CALIFA face-on spiral galaxies}",
      journal = {\aap},
         year = 2016,
        month = mar,
       volume = {587},
          eid = {A70},
        pages = {A70},
          doi = {10.1051/0004-6361/201527450},
archivePrefix = {arXiv},
       eprint = {1601.01542},
 primaryClass = {astro-ph.GA},
       adsurl = {https://ui.adsabs.harvard.edu/abs/2016A&A...587A..70S}
}

@ARTICLE{Belfiore_2017MNRAS.469..151B,
       author = {{Belfiore}, Francesco and {Maiolino}, Roberto and {Tremonti}, Christy and {S{\'a}nchez}, Sebastian F. and {Bundy}, Kevin and {Bershady}, Matthew and {Westfall}, Kyle and {Lin}, Lihwai and {Drory}, Niv and {Boquien}, M{\'e}d{\'e}ric and {Thomas}, Daniel and {Brinkmann}, Jonathan},
        title = "{SDSS IV MaNGA - metallicity and nitrogen abundance gradients in local galaxies}",
      journal = {\mnras},
         year = 2017,
        month = jul,
       volume = {469},
       number = {1},
        pages = {151-170},
          doi = {10.1093/mnras/stx789},
archivePrefix = {arXiv},
       eprint = {1703.03813},
 primaryClass = {astro-ph.GA},
       adsurl = {https://ui.adsabs.harvard.edu/abs/2017MNRAS.469..151B}
}

@ARTICLE{Werk_2011ApJ...735...71W,
       author = {{Werk}, J.~K. and {Putman}, M.~E. and {Meurer}, G.~R. and {Santiago-Figueroa}, N.},
        title = "{Metal Transport to the Gaseous Outskirts of Galaxies}",
      journal = {\apj},
         year = 2011,
        month = jul,
       volume = {735},
       number = {2},
          eid = {71},
        pages = {71},
          doi = {10.1088/0004-637X/735/2/71},
archivePrefix = {arXiv},
       eprint = {1104.3897},
 primaryClass = {astro-ph.CO},
       adsurl = {https://ui.adsabs.harvard.edu/abs/2011ApJ...735...71W}
}

@ARTICLE{Bresolin_2012ApJ...750..122B,
       author = {{Bresolin}, Fabio and {Kennicutt}, Robert C. and {Ryan-Weber}, Emma},
        title = "{Gas Metallicities in the Extended Disks of NGC 1512 and NGC 3621. Chemical Signatures of Metal Mixing or Enriched Gas Accretion?}",
      journal = {\apj},
         year = 2012,
        month = may,
       volume = {750},
       number = {2},
          eid = {122},
        pages = {122},
          doi = {10.1088/0004-637X/750/2/122},
archivePrefix = {arXiv},
       eprint = {1203.0956},
 primaryClass = {astro-ph.CO},
       adsurl = {https://ui.adsabs.harvard.edu/abs/2012ApJ...750..122B}
}

@ARTICLE{Kewley_2010ApJ...721L..48K,
       author = {{Kewley}, Lisa J. and {Rupke}, David and {Zahid}, H. Jabran and {Geller}, Margaret J. and {Barton}, Elizabeth J.},
        title = "{Metallicity Gradients and Gas Flows in Galaxy Pairs}",
      journal = {\apjl},
         year = 2010,
        month = sep,
       volume = {721},
       number = {1},
        pages = {L48-L52},
          doi = {10.1088/2041-8205/721/1/L48},
archivePrefix = {arXiv},
       eprint = {1008.2204},
 primaryClass = {astro-ph.CO},
       adsurl = {https://ui.adsabs.harvard.edu/abs/2010ApJ...721L..48K}
}

@ARTICLE{Gim_2021ApJ...922...69G,
       author = {{Gim}, Hansung B. and {Borthakur}, Sanchayeeta and {Momjian}, Emmanuel and {Padave}, Mansi and {Jansen}, Rolf A. and {Nelson}, Dylan and {Heckman}, Timothy M. and {Kennicutt}, Jr., Robert C. and {Fox}, Andrew J. and {Pineda}, Jorge L. and {Thilker}, David and {Kauffmann}, Guinevere and {Tumlinson}, Jason},
        title = "{DIISC-I: The Discovery of Kinematically Anomalous H I Clouds in M 100}",
      journal = {\apj},
         year = 2021,
        month = nov,
       volume = {922},
       number = {1},
          eid = {69},
        pages = {69},
          doi = {10.3847/1538-4357/ac2303},
archivePrefix = {arXiv},
       eprint = {2109.06247},
 primaryClass = {astro-ph.GA},
       adsurl = {https://ui.adsabs.harvard.edu/abs/2021ApJ...922...69G}
}

@ARTICLE{Martin_1994ApJ...424..599M,
       author = {{Martin}, Pierre and {Roy}, Jean-Rene},
        title = "{The Influence of Bars on the Chemical Composition of Spiral Galaxies}",
      journal = {\apj},
         year = 1994,
        month = apr,
       volume = {424},
        pages = {599},
          doi = {10.1086/173917},
       adsurl = {https://ui.adsabs.harvard.edu/abs/1994ApJ...424..599M}
}

@ARTICLE{Sanchez-Menguiano_2026MNRAS.545f2029S,
       author = {{S{\'a}nchez-Menguiano}, Laura and {Gadotti}, Dimitri A. and {Zurita}, Almudena and {Florido}, Estrella and {P{\'e}rez}, Isabel and {Coelho}, Paula and {Falc{\'o}n-Barroso}, Jes{\'u}s and {Kim}, Taehyun and {de Lorenzo-C{\'a}ceres}, Adriana and {Lugo-Aranda}, Alejandra Z. and {Neumann}, Justus and {de S{\'a}-Freitas}, Camila and {S{\'a}nchez-Bl{\'a}zquez}, Patricia},
        title = "{The impact of bars on the properties of H II regions in the TIMER survey}",
      journal = {\mnras},
         year = 2026,
        month = jan,
       volume = {545},
       number = {1},
          eid = {staf2029},
        pages = {staf2029},
          doi = {10.1093/mnras/staf2029},
archivePrefix = {arXiv},
       eprint = {2512.15637},
 primaryClass = {astro-ph.GA},
       adsurl = {https://ui.adsabs.harvard.edu/abs/2026MNRAS.545f2029S}
}

@ARTICLE{Pilyugin_2023A&A...676A..57P,
       author = {{Pilyugin}, L.~S. and {Tautvai{\v{s}}ien{\.{e}}}, G. and {Lara-L{\'o}pez}, M.~A.},
        title = "{Searching for Milky Way twins: Radial abundance distribution as a strict criterion}",
      journal = {\aap},
         year = 2023,
        month = aug,
       volume = {676},
          eid = {A57},
        pages = {A57},
          doi = {10.1051/0004-6361/202346503},
archivePrefix = {arXiv},
       eprint = {2306.09854},
 primaryClass = {astro-ph.GA},
       adsurl = {https://ui.adsabs.harvard.edu/abs/2023A&A...676A..57P}
}

@ARTICLE{Hwang_2019ApJ...872..144H,
       author = {{Hwang}, Hsiang-Chih and {Barrera-Ballesteros}, Jorge K. and {Heckman}, Timothy M. and {Rowlands}, Kate and {Lin}, Lihwai and {Rodriguez-Gomez}, Vicente and {Pan}, Hsi-An and {Hsieh}, Bau-Ching and {S{\'a}nchez}, Sebastian and {Bizyaev}, Dmitry and {S{\'a}nchez Almeida}, Jorge and {Thilker}, David A. and {Lotz}, Jennifer M. and {Jones}, Amy and {Nair}, Preethi and {Andrews}, Brett H. and {Drory}, Niv},
        title = "{Anomalously Low-metallicity Regions in MaNGA Star-forming Galaxies: Accretion Caught in Action?}",
      journal = {\apj},
         year = 2019,
        month = feb,
       volume = {872},
       number = {2},
          eid = {144},
        pages = {144},
          doi = {10.3847/1538-4357/aaf7a3},
archivePrefix = {arXiv},
       eprint = {1812.04614},
 primaryClass = {astro-ph.GA},
       adsurl = {https://ui.adsabs.harvard.edu/abs/2019ApJ...872..144H}
}

@software{Bradley_larry_bradley_2023_7946442,
  author       = {Larry Bradley},
  title        = {astropy/photutils: 1.8.0},
  month        = may,
  year         = 2023,
  publisher    = {Zenodo},
  version      = {1.8.0},
  doi          = {10.5281/zenodo.7946442},
  url          = {https://doi.org/10.5281/zenodo.7946442},
}

@ARTICLE{Schinnerer_2019ApJ...887...49S,
       author = {{Schinnerer}, Eva and {Hughes}, Annie and {Leroy}, Adam and {Groves}, Brent and {Blanc}, Guillermo A. and {Kreckel}, Kathryn and {Bigiel}, Frank and {Chevance}, M{\'e}lanie and {Dale}, Daniel and {Emsellem}, Eric and {Faesi}, Christopher and {Glover}, Simon and {Grasha}, Kathryn and {Henshaw}, Jonathan and {Hygate}, Alexander and {Kruijssen}, J.~M. Diederik and {Meidt}, Sharon and {Pety}, Jerome and {Querejeta}, Miguel and {Rosolowsky}, Erik and {Saito}, Toshiki and {Schruba}, Andreas and {Sun}, Jiayi and {Utomo}, Dyas},
        title = "{The Gas-Star Formation Cycle in Nearby Star-forming Galaxies. I. Assessment of Multi-scale Variations}",
      journal = {\apj},
         year = 2019,
        month = dec,
       volume = {887},
       number = {1},
          eid = {49},
        pages = {49},
          doi = {10.3847/1538-4357/ab50c2},
archivePrefix = {arXiv},
       eprint = {1910.10520},
 primaryClass = {astro-ph.GA},
       adsurl = {https://ui.adsabs.harvard.edu/abs/2019ApJ...887...49S}
}

@ARTICLE{Semenov_2021ApJ...918...13S,
       author = {{Semenov}, Vadim A. and {Kravtsov}, Andrey V. and {Gnedin}, Nickolay Y.},
        title = "{Spatial Decorrelation of Young Stars and Dense Gas as a Probe of the Star Formation-Feedback Cycle in Galaxies}",
      journal = {\apj},
         year = 2021,
        month = sep,
       volume = {918},
       number = {1},
          eid = {13},
        pages = {13},
          doi = {10.3847/1538-4357/ac0a77},
archivePrefix = {arXiv},
       eprint = {2103.13406},
 primaryClass = {astro-ph.GA},
       adsurl = {https://ui.adsabs.harvard.edu/abs/2021ApJ...918...13S}
}

@ARTICLE{Kreckel_2019ApJ...887...80K,
       author = {{Kreckel}, K. and {Ho}, I.-T. and {Blanc}, G.~A. and {Groves}, B. and {Santoro}, F. and {Schinnerer}, E. and {Bigiel}, F. and {Chevance}, M. and {Congiu}, E. and {Emsellem}, E. and {Faesi}, C. and {Glover}, S.~C.~O. and {Grasha}, K. and {Kruijssen}, J.~M.~D. and {Lang}, P. and {Leroy}, A.~K. and {Meidt}, S.~E. and {McElroy}, R. and {Pety}, J. and {Rosolowsky}, E. and {Saito}, T. and {Sandstrom}, K. and {Sanchez-Blazquez}, P. and {Schruba}, A.},
        title = "{Mapping Metallicity Variations across Nearby Galaxy Disks}",
      journal = {\apj},
         year = 2019,
        month = dec,
       volume = {887},
       number = {1},
          eid = {80},
        pages = {80},
          doi = {10.3847/1538-4357/ab5115},
archivePrefix = {arXiv},
       eprint = {1910.07190},
 primaryClass = {astro-ph.GA},
       adsurl = {https://ui.adsabs.harvard.edu/abs/2019ApJ...887...80K}
}

@ARTICLE{Kewley_2008ApJ...681.1183K,
       author = {{Kewley}, Lisa J. and {Ellison}, Sara L.},
        title = "{Metallicity Calibrations and the Mass-Metallicity Relation for Star-forming Galaxies}",
      journal = {\apj},
         year = 2008,
        month = jul,
       volume = {681},
       number = {2},
        pages = {1183-1204},
          doi = {10.1086/587500},
archivePrefix = {arXiv},
       eprint = {0801.1849},
 primaryClass = {astro-ph},
       adsurl = {https://ui.adsabs.harvard.edu/abs/2008ApJ...681.1183K}
}

@ARTICLE{Poetrodjojo_2019MNRAS.487...79P,
       author = {{Poetrodjojo}, Henry and {D'Agostino}, Joshua J. and {Groves}, Brent and {Kewley}, Lisa and {Ho}, I.-Ting and {Rich}, Jeff and {Madore}, Barry F. and {Seibert}, Mark},
        title = "{The effects of diffuse ionized gas and spatial resolution on metallicity gradients: TYPHOON two-dimensional spectrophotometry of M83}",
      journal = {\mnras},
         year = 2019,
        month = jul,
       volume = {487},
       number = {1},
        pages = {79-96},
          doi = {10.1093/mnras/stz1241},
archivePrefix = {arXiv},
       eprint = {1905.03251},
 primaryClass = {astro-ph.GA},
       adsurl = {https://ui.adsabs.harvard.edu/abs/2019MNRAS.487...79P}
}

@ARTICLE{Pilyugin_2016MNRAS.457.3678P,
       author = {{Pilyugin}, L.~S. and {Grebel}, E.~K.},
        title = "{New calibrations for abundance determinations in H II regions}",
      journal = {\mnras},
         year = 2016,
        month = apr,
       volume = {457},
       number = {4},
        pages = {3678-3692},
          doi = {10.1093/mnras/stw238},
archivePrefix = {arXiv},
       eprint = {1601.08217},
 primaryClass = {astro-ph.GA},
       adsurl = {https://ui.adsabs.harvard.edu/abs/2016MNRAS.457.3678P}
}

@article{YangKrumholz2012,
  author  = {Yang, Chao-Chin and Krumholz, Mark R.},
  title   = {Thermal-instability-driven Turbulent Mixing in
             Galactic Disks. I. Effective Mixing of Metals},
  journal = {The Astrophysical Journal},
  year    = {2012},
  volume  = {758},
  number  = {1},
  pages   = {48},
  doi     = {10.1088/0004-637X/758/1/48}
}

@article{Petit2015,
  author  = {Petit, Antoine C. and Krumholz, Mark R.
             and Goldbaum, Nathan J. and Forbes, John C.},
  title   = {Mixing and Transport of Metals by Gravitational
             Instability-driven Turbulence in Galactic Discs},
  journal = {Monthly Notices of the Royal Astronomical Society},
  year    = {2015},
  volume  = {449},
  number  = {3},
  pages   = {2588--2597},
  doi     = {10.1093/mnras/stv493}
}

@article{Lilly2013,
  author  = {Lilly, Simon J. and Carollo, C. Marcella
             and Pipino, Antonio and Renzini, Alvio
             and Peng, Ying-jie},
  title   = {Gas Regulation of Galaxies: The Evolution of the
             Cosmic Specific Star Formation Rate, the
             Metallicity--Mass--Star-formation Rate Relation,
             and the Stellar Content of Halos},
  journal = {The Astrophysical Journal},
  year    = {2013},
  volume  = {772},
  number  = {2},
  pages   = {119},
  doi     = {10.1088/0004-637X/772/2/119}
}

@ARTICLE{Putman_2012ARA&A..50..491P,
       author = {{Putman}, M.~E. and {Peek}, J.~E.~G. and {Joung}, M.~R.},
        title = "{Gaseous Galaxy Halos}",
      journal = {\araa},
         year = 2012,
        month = sep,
       volume = {50},
        pages = {491-529},
          doi = {10.1146/annurev-astro-081811-125612},
archivePrefix = {arXiv},
       eprint = {1207.4837},
 primaryClass = {astro-ph.GA},
       adsurl = {https://ui.adsabs.harvard.edu/abs/2012ARA&A..50..491P}
}

@ARTICLE{Lucchini_2025ApJ...990..118L,
       author = {{Lucchini}, Scott and {Han}, Jiwon Jesse and {Hernquist}, Lars and {Conroy}, Charlie and {Fox}, Andrew J.},
        title = "{Invisible Accretion: Ionized Envelopes of TNG50 HVCs Can Sustain Star Formation}",
      journal = {\apj},
         year = 2025,
        month = sep,
       volume = {990},
       number = {2},
          eid = {118},
        pages = {118},
          doi = {10.3847/1538-4357/adf3b3},
archivePrefix = {arXiv},
       eprint = {2507.18687},
 primaryClass = {astro-ph.GA},
       adsurl = {https://ui.adsabs.harvard.edu/abs/2025ApJ...990..118L}
}

@ARTICLE{Ceverino_2016MNRAS.457.2605C,
       author = {{Ceverino}, Daniel and {S{\'a}nchez Almeida}, Jorge and {Mu{\~n}oz Tu{\~n}{\'o}n}, Casiana and {Dekel}, Avishai and {Elmegreen}, Bruce G. and {Elmegreen}, Debra M. and {Primack}, Joel},
        title = "{Gas inflow and metallicity drops in star-forming galaxies}",
      journal = {\mnras},
         year = 2016,
        month = apr,
       volume = {457},
       number = {3},
        pages = {2605-2612},
          doi = {10.1093/mnras/stw064},
archivePrefix = {arXiv},
       eprint = {1509.02051},
 primaryClass = {astro-ph.GA},
       adsurl = {https://ui.adsabs.harvard.edu/abs/2016MNRAS.457.2605C}
}

@article{astropy:2013,
Adsurl = {http://adsabs.harvard.edu/abs/2013A%26A...558A..33A},
Archiveprefix = {arXiv},
Author = {{Astropy Collaboration} and {Robitaille}, T.~P. and {Tollerud}, E.~J. and {Greenfield}, P. and {Droettboom}, M. and {Bray}, E. and {Aldcroft}, T. and {Davis}, M. and {Ginsburg}, A. and {Price-Whelan}, A.~M. and {Kerzendorf}, W.~E. and {Conley}, A. and {Crighton}, N. and {Barbary}, K. and {Muna}, D. and {Ferguson}, H. and {Grollier}, F. and {Parikh}, M.~M. and {Nair}, P.~H. and {Unther}, H.~M. and {Deil}, C. and {Woillez}, J. and {Conseil}, S. and {Kramer}, R. and {Turner}, J.~E.~H. and {Singer}, L. and {Fox}, R. and {Weaver}, B.~A. and {Zabalza}, V. and {Edwards}, Z.~I. and {Azalee Bostroem}, K. and {Burke}, D.~J. and {Casey}, A.~R. and {Crawford}, S.~M. and {Dencheva}, N. and {Ely}, J. and {Jenness}, T. and {Labrie}, K. and {Lim}, P.~L. and {Pierfederici}, F. and {Pontzen}, A. and {Ptak}, A. and {Refsdal}, B. and {Servillat}, M. and {Streicher}, O.},
Doi = {10.1051/0004-6361/201322068},
Eid = {A33},
Eprint = {1307.6212},
Journal = {\aap},
Month = oct,
Pages = {A33},
Primaryclass = {astro-ph.IM},
Title = {{Astropy: A community Python package for astronomy}},
Volume = 558,
Year = 2013}

@ARTICLE{astropy:2018,
       author = {{Astropy Collaboration} and {Price-Whelan}, A.~M. and
         {Sip{\H{o}}cz}, B.~M. and {G{\"u}nther}, H.~M. and {Lim}, P.~L. and
         {Crawford}, S.~M. and {Conseil}, S. and {Shupe}, D.~L. and
         {Craig}, M.~W. and {Dencheva}, N. and {Ginsburg}, A. and {Vand
        erPlas}, J.~T. and {Bradley}, L.~D. and {P{\'e}rez-Su{\'a}rez}, D. and
         {de Val-Borro}, M. and {Aldcroft}, T.~L. and {Cruz}, K.~L. and
         {Robitaille}, T.~P. and {Tollerud}, E.~J. and {Ardelean}, C. and
         {Babej}, T. and {Bach}, Y.~P. and {Bachetti}, M. and {Bakanov}, A.~V. and
         {Bamford}, S.~P. and {Barentsen}, G. and {Barmby}, P. and
         {Baumbach}, A. and {Berry}, K.~L. and {Biscani}, F. and {Boquien}, M. and
         {Bostroem}, K.~A. and {Bouma}, L.~G. and {Brammer}, G.~B. and
         {Bray}, E.~M. and {Breytenbach}, H. and {Buddelmeijer}, H. and
         {Burke}, D.~J. and {Calderone}, G. and {Cano Rodr{\'\i}guez}, J.~L. and
         {Cara}, M. and {Cardoso}, J.~V.~M. and {Cheedella}, S. and {Copin}, Y. and
         {Corrales}, L. and {Crichton}, D. and {D'Avella}, D. and {Deil}, C. and
         {Depagne}, {\'E}. and {Dietrich}, J.~P. and {Donath}, A. and
         {Droettboom}, M. and {Earl}, N. and {Erben}, T. and {Fabbro}, S. and
         {Ferreira}, L.~A. and {Finethy}, T. and {Fox}, R.~T. and
         {Garrison}, L.~H. and {Gibbons}, S.~L.~J. and {Goldstein}, D.~A. and
         {Gommers}, R. and {Greco}, J.~P. and {Greenfield}, P. and
         {Groener}, A.~M. and {Grollier}, F. and {Hagen}, A. and {Hirst}, P. and
         {Homeier}, D. and {Horton}, A.~J. and {Hosseinzadeh}, G. and {Hu}, L. and
         {Hunkeler}, J.~S. and {Ivezi{\'c}}, {\v{Z}}. and {Jain}, A. and
         {Jenness}, T. and {Kanarek}, G. and {Kendrew}, S. and {Kern}, N.~S. and
         {Kerzendorf}, W.~E. and {Khvalko}, A. and {King}, J. and {Kirkby}, D. and
         {Kulkarni}, A.~M. and {Kumar}, A. and {Lee}, A. and {Lenz}, D. and
         {Littlefair}, S.~P. and {Ma}, Z. and {Macleod}, D.~M. and
         {Mastropietro}, M. and {McCully}, C. and {Montagnac}, S. and
         {Morris}, B.~M. and {Mueller}, M. and {Mumford}, S.~J. and {Muna}, D. and
         {Murphy}, N.~A. and {Nelson}, S. and {Nguyen}, G.~H. and
         {Ninan}, J.~P. and {N{\"o}the}, M. and {Ogaz}, S. and {Oh}, S. and
         {Parejko}, J.~K. and {Parley}, N. and {Pascual}, S. and {Patil}, R. and
         {Patil}, A.~A. and {Plunkett}, A.~L. and {Prochaska}, J.~X. and
         {Rastogi}, T. and {Reddy Janga}, V. and {Sabater}, J. and
         {Sakurikar}, P. and {Seifert}, M. and {Sherbert}, L.~E. and
         {Sherwood-Taylor}, H. and {Shih}, A.~Y. and {Sick}, J. and
         {Silbiger}, M.~T. and {Singanamalla}, S. and {Singer}, L.~P. and
         {Sladen}, P.~H. and {Sooley}, K.~A. and {Sornarajah}, S. and
         {Streicher}, O. and {Teuben}, P. and {Thomas}, S.~W. and
         {Tremblay}, G.~R. and {Turner}, J.~E.~H. and {Terr{\'o}n}, V. and
         {van Kerkwijk}, M.~H. and {de la Vega}, A. and {Watkins}, L.~L. and
         {Weaver}, B.~A. and {Whitmore}, J.~B. and {Woillez}, J. and
         {Zabalza}, V. and {Astropy Contributors}},
        title = "{The Astropy Project: Building an Open-science Project and Status of the v2.0 Core Package}",
      journal = {\aj},
         year = 2018,
        month = sep,
       volume = {156},
       number = {3},
          eid = {123},
        pages = {123},
          doi = {10.3847/1538-3881/aabc4f},
archivePrefix = {arXiv},
       eprint = {1801.02634},
 primaryClass = {astro-ph.IM},
       adsurl = {https://ui.adsabs.harvard.edu/abs/2018AJ....156..123A}
}

@ARTICLE{astropy:2022,
       author = {{Astropy Collaboration} and {Price-Whelan}, Adrian M. and {Lim}, Pey Lian and {Earl}, Nicholas and {Starkman}, Nathaniel and {Bradley}, Larry and {Shupe}, David L. and {Patil}, Aarya A. and {Corrales}, Lia and {Brasseur}, C.~E. and {N{"o}the}, Maximilian and {Donath}, Axel and {Tollerud}, Erik and {Morris}, Brett M. and {Ginsburg}, Adam and {Vaher}, Eero and {Weaver}, Benjamin A. and {Tocknell}, James and {Jamieson}, William and {van Kerkwijk}, Marten H. and {Robitaille}, Thomas P. and {Merry}, Bruce and {Bachetti}, Matteo and {G{"u}nther}, H. Moritz and {Aldcroft}, Thomas L. and {Alvarado-Montes}, Jaime A. and {Archibald}, Anne M. and {B{'o}di}, Attila and {Bapat}, Shreyas and {Barentsen}, Geert and {Baz{'a}n}, Juanjo and {Biswas}, Manish and {Boquien}, M{'e}d{'e}ric and {Burke}, D.~J. and {Cara}, Daria and {Cara}, Mihai and {Conroy}, Kyle E. and {Conseil}, Simon and {Craig}, Matthew W. and {Cross}, Robert M. and {Cruz}, Kelle L. and {D'Eugenio}, Francesco and {Dencheva}, Nadia and {Devillepoix}, Hadrien A.~R. and {Dietrich}, J{"o}rg P. and {Eigenbrot}, Arthur Davis and {Erben}, Thomas and {Ferreira}, Leonardo and {Foreman-Mackey}, Daniel and {Fox}, Ryan and {Freij}, Nabil and {Garg}, Suyog and {Geda}, Robel and {Glattly}, Lauren and {Gondhalekar}, Yash and {Gordon}, Karl D. and {Grant}, David and {Greenfield}, Perry and {Groener}, Austen M. and {Guest}, Steve and {Gurovich}, Sebastian and {Handberg}, Rasmus and {Hart}, Akeem and {Hatfield-Dodds}, Zac and {Homeier}, Derek and {Hosseinzadeh}, Griffin and {Jenness}, Tim and {Jones}, Craig K. and {Joseph}, Prajwel and {Kalmbach}, J. Bryce and {Karamehmetoglu}, Emir and {Ka{l}uszy{'n}ski}, Miko{l}aj and {Kelley}, Michael S.~P. and {Kern}, Nicholas and {Kerzendorf}, Wolfgang E. and {Koch}, Eric W. and {Kulumani}, Shankar and {Lee}, Antony and {Ly}, Chun and {Ma}, Zhiyuan and {MacBride}, Conor and {Maljaars}, Jakob M. and {Muna}, Demitri and {Murphy}, N.~A. and {Norman}, Henrik and {O'Steen}, Richard and {Oman}, Kyle A. and {Pacifici}, Camilla and {Pascual}, Sergio and {Pascual-Granado}, J. and {Patil}, Rohit R. and {Perren}, Gabriel I. and {Pickering}, Timothy E. and {Rastogi}, Tanuj and {Roulston}, Benjamin R. and {Ryan}, Daniel F. and {Rykoff}, Eli S. and {Sabater}, Jose and {Sakurikar}, Parikshit and {Salgado}, Jes{'u}s and {Sanghi}, Aniket and {Saunders}, Nicholas and {Savchenko}, Volodymyr and {Schwardt}, Ludwig and {Seifert-Eckert}, Michael and {Shih}, Albert Y. and {Jain}, Anany Shrey and {Shukla}, Gyanendra and {Sick}, Jonathan and {Simpson}, Chris and {Singanamalla}, Sudheesh and {Singer}, Leo P. and {Singhal}, Jaladh and {Sinha}, Manodeep and {Sip{H{o}}cz}, Brigitta M. and {Spitler}, Lee R. and {Stansby}, David and {Streicher}, Ole and {{{S}}umak}, Jani and {Swinbank}, John D. and {Taranu}, Dan S. and {Tewary}, Nikita and {Tremblay}, Grant R. and {Val-Borro}, Miguel de and {Van Kooten}, Samuel J. and {Vasovi{'c}}, Zlatan and {Verma}, Shresth and {de Miranda Cardoso}, Jos{'e} Vin{'i}cius and {Williams}, Peter K.~G. and {Wilson}, Tom J. and {Winkel}, Benjamin and {Wood-Vasey}, W.~M. and {Xue}, Rui and {Yoachim}, Peter and {Zhang}, Chen and {Zonca}, Andrea and {Astropy Project Contributors}},
        title = "{The Astropy Project: Sustaining and Growing a Community-oriented Open-source Project and the Latest Major Release (v5.0) of the Core Package}",
      journal = {\apj},
         year = 2022,
        month = aug,
       volume = {935},
       number = {2},
          eid = {167},
        pages = {167},
          doi = {10.3847/1538-4357/ac7c74},
archivePrefix = {arXiv},
       eprint = {2206.14220},
 primaryClass = {astro-ph.IM},
       adsurl = {https://ui.adsabs.harvard.edu/abs/2022ApJ...935..167A}
}

@ARTICLE{Perez_2007CSE.....9c..21P,
       author = {{Perez}, Fernando and {Granger}, Brian E.},
        title = "{IPython: A System for Interactive Scientific Computing}",
      journal = {Computing in Science and Engineering},
         year = 2007,
        month = jan,
       volume = {9},
       number = {3},
        pages = {21-29},
          doi = {10.1109/MCSE.2007.53},
       adsurl = {https://ui.adsabs.harvard.edu/abs/2007CSE.....9c..21P}
}

@INCOLLECTION{Kluyver_2016ppap.book...87K,
       author = {{Kluyver}, Thomas and {Ragan-Kelley}, Benjain and {P{\'e}rez}, Fernando and {Granger}, Brian and {Bussonnier}, Matthias and {Frederic}, Jonathan and {Kelley}, Kyle and {Hamrick}, Jessica and {Grout}, Jason and {Corlay}, Sylvain and {Ivanov}, Paul and {Avila}, Dami{\'a}n and {Abdalla}, Safia and {Willing}, Carol and {Jupyter Development Team}},
        title = "{Jupyter Notebooks{\textemdash}a publishing format for reproducible computational workflows}",
    booktitle = {IOS Press},
         year = 2016,
        pages = {87-90},
          doi = {10.3233/978-1-61499-649-1-87},
       adsurl = {https://ui.adsabs.harvard.edu/abs/2016ppap.book...87K}
}

@Article{Hunter:2007,
  Author    = {Hunter, J. D.},
  Title     = {Matplotlib: A 2D graphics environment},
  Journal   = {Computing in Science \& Engineering},
  Volume    = {9},
  Number    = {3},
  Pages     = {90--95},
  publisher = {IEEE COMPUTER SOC},
  doi       = {10.1109/MCSE.2007.55},
  year      = 2007
}

@Article{         harris2020array,
 title         = {Array programming with {NumPy}},
 author        = {Charles R. Harris and K. Jarrod Millman and St{\'{e}}fan J.
                 van der Walt and Ralf Gommers and Pauli Virtanen and David
                 Cournapeau and Eric Wieser and Julian Taylor and Sebastian
                 Berg and Nathaniel J. Smith and Robert Kern and Matti Picus
                 and Stephan Hoyer and Marten H. van Kerkwijk and Matthew
                 Brett and Allan Haldane and Jaime Fern{\'{a}}ndez del
                 R{\'{i}}o and Mark Wiebe and Pearu Peterson and Pierre
                 G{\'{e}}rard-Marchant and Kevin Sheppard and Tyler Reddy and
                 Warren Weckesser and Hameer Abbasi and Christoph Gohlke and
                 Travis E. Oliphant},
 year          = {2020},
 month         = sep,
 journal       = {Nature},
 volume        = {585},
 number        = {7825},
 pages         = {357--362},
 doi           = {10.1038/s41586-020-2649-2},
 publisher     = {Springer Science and Business Media {LLC}},
 url           = {https://doi.org/10.1038/s41586-020-2649-2}
}

@MISC{specutils,
       author = {{Earl}, Nicholas and {Tollerud}, Erik and {O'Steen}, Ricky and {Brechmos} and {Kerzendorf}, Wolfgang and {Busko}, Ivo and {Shaileshahuja} and {D'Avella}, Dan and {Robitaille}, Thomas and {Ginsburg}, Adam and {Homeier}, Derek and {Sip{\H{o}}cz}, Brigitta and {Averbukh}, Jesse and {Lim}, P.~L. and {Tocknell}, James and {Cherinka}, Brian and {Ogaz}, Sara and {Geda}, Robel and {Davies}, James and {G{\"u}nther}, Hans Moritz and {Conroy}, Kyle and {Barbary}, Kyle and {Foster}, Jonathan and {Droettboom}, Michael and {Torres}, Simon and {Bray}, E.~M. and {Casey}, Andy and {Teuben}, Peter and {Crawford}, Steve and {Ferguson}, Henry},
        title = "{astropy/specutils: V1.8.0}",
 howpublished = {Zenodo},
         year = 2022,
        month = aug,
          eid = {10.5281/zenodo.7015214},
          doi = {10.5281/zenodo.7015214},
      version = {v1.8.0},
    publisher = {Zenodo},
       adsurl = {https://ui.adsabs.harvard.edu/abs/2022zndo...7015214E}
}

@article{borthakur2016,
  author  = {Borthakur, Sanchayeeta and Heckman, Timothy and
             Tumlinson, Jason and Bordoloi, Rongmon and
             Kauffmann, Guinevere and Catinella, Barbara and
             Schiminovich, David and Dav{\\'e}, Romeel and
             Moran, Sean M. and Saintonge, Am{\\'e}lie},
  title   = {The Properties of the Circumgalactic Medium in Red and
             Blue Galaxies: Results from the COS-GASS+COS-Halos Surveys},
  journal = {The Astrophysical Journal},
  year    = {2016},
  volume  = {833},
  number  = {2},
  pages   = {259},
  doi     = {10.3847/1538-4357/833/2/259}
}

@article{borthakur2024diisc,
  author        = {Borthakur, Sanchayeeta and Padave, Mansi and
                   Heckman, Timothy and Gim, Hansung B. and
                   Olvera, Alejandro J. and Koplitz, Brad and
                   Momjian, Emmanuel and Jansen, Rolf A. and
                   Thilker, David and Kauffmann, Guinevere and
                   Fox, Andrew J. and Tumlinson, Jason and
                   Kennicutt, Robert C. and Nelson, Dylan and
                   Monckiewicz, Jacqueline and Naab, Thorsten},
  title         = {{DIISC Survey}: Deciphering the Interplay Between
                   the Interstellar Medium, Stars, and the
                   Circumgalactic Medium Survey},
  journal       = {arXiv e-prints},
  year          = {2024},
  eid           = {arXiv:2409.12554},
  archivePrefix = {arXiv},
  eprint        = {2409.12554},
  primaryClass  = {astro-ph.GA},
  doi           = {10.48550/arXiv.2409.12554}
}

@article{Keres2005,
  author = {Kere\v{s}, Du\v{s}an and Katz, Neal and Weinberg, David H. and Dav\'{e}, Romeel},
  title = {How Do Galaxies Get Their Gas?},
  journal = {MNRAS},
  year = {2005},
  volume = {363},
  pages = {2--28},
  doi = {10.1111/j.1365-2966.2005.09451.x}
}

@article{FaucherGiguereOh2023,
  author = {Faucher-Gigu\`ere, Claude-Andr\'{e} and Oh, S. Peng},
  title = {Key Physical Processes in the Circumgalactic Medium},
  journal = {ARA\&A},
  year = {2023},
  volume = {61},
  pages = {131--195},
  doi = {10.1146/annurev-astro-052920-125203}
}

@article{Sancisi2008,
  author = {Sancisi, Renzo and Fraternali, Filippo and Oosterloo, Tom and van der Hulst, Thijs},
  title = {Cold Gas Accretion in Galaxies},
  journal = {A\&A Review},
  year = {2008},
  volume = {15},
  pages = {189--223},
  doi = {10.1007/s00159-008-0010-0}
}

@article{DiTeodoroPeek2021,
  author = {Di Teodoro, Enrico M. and Peek, J. E. G.},
  title = {Radial Motions and Radial Gas Flows in Local Spiral Galaxies},
  journal = {ApJ},
  year = {2021},
  volume = {923},
  pages = {220},
  doi = {10.3847/1538-4357/ac2cbd}
}

@article{FinlatorDave2008,
  author = {Finlator, Kristian and Dav\'{e}, Romeel},
  title = {The Origin of the Galaxy Mass--Metallicity Relation and Implications for Galactic Outflows},
  journal = {MNRAS},
  year = {2008},
  volume = {385},
  pages = {2181--2204},
  doi = {10.1111/j.1365-2966.2008.12991.x}
}

@ARTICLE{Olvera2026arXiv260926884O,
       author = {{Olvera}, Alejandro J. and {Borthakur}, Sanchayeeta and {Di Teodoro}, Enrico M. and {Saha}, Kanak and {Padave}, Mansi and {Gim}, Hansung B. and {Momjian}, Emmanuel},
        title = "{DIISC-VII: Linking Radial Gas Motions to Anomalously Low Metallicity Star-forming Regions in the XUV Disk of NGC 3344}",
      journal = {arXiv e-prints},
         year = 2026,
        month = sep,
          eid = {arXiv:2609.26884},
        pages = {arXiv:2609.26884},
archivePrefix = {arXiv},
       eprint = {2609.26884},
 primaryClass = {astro-ph.GA},
       adsurl = {https://ui.adsabs.harvard.edu/abs/2026arXiv260926884O}
}

@ARTICLE{Yao2022A&A...661A.112Y,
       author = {{Yao}, Yao and {Chen}, Guangwen and {Liu}, Haiyang and {Chen}, Xinkai and {Lin}, Zesen and {Zhang}, Hong-Xin and {Gao}, Yulong and {Kong}, Xu},
        title = "{Spatially resolved mass-metallicity relation at z {\ensuremath{\sim}} 0.26 from the MUSE-Wide Survey}",
      journal = {\aap},
         year = 2022,
        month = may,
       volume = {661},
          eid = {A112},
        pages = {A112},
          doi = {10.1051/0004-6361/202243104},
archivePrefix = {arXiv},
       eprint = {2203.13681},
 primaryClass = {astro-ph.GA},
       adsurl = {https://ui.adsabs.harvard.edu/abs/2022A&A...661A.112Y}
}

@ARTICLE{Wong_2009ApJ...705..650W,
       author = {{Wong}, Tony},
        title = "{On the Timescale for Star Formation in Galaxies}",
      journal = {\apj},
         year = 2009,
        month = nov,
       volume = {705},
       number = {1},
        pages = {650-658},
          doi = {10.1088/0004-637X/705/1/650},
archivePrefix = {arXiv},
       eprint = {0909.2243},
 primaryClass = {astro-ph.GA},
       adsurl = {https://ui.adsabs.harvard.edu/abs/2009ApJ...705..650W}
}

@article{Wakker1999LowMetallicityAccretion,
  author  = {Wakker, B. P. and Howk, J. C. and Savage, B. D. and
             van Woerden, H. and Tufte, S. L. and Schwarz, U. J. and
             Benjamin, R. and Reynolds, R. J. and Peletier, R. F.},
  title   = {Accretion of Low-Metallicity Gas by the Milky Way},
  journal = {Nature},
  year    = {1999},
  volume  = {402},
  pages   = {388--390},
  doi     = {10.1038/46598}
}

@article{Wakker2001DistancesMetallicities,
  author  = {Wakker, Bart P.},
  title   = {Distances and Metallicities of High- and
             Intermediate-Velocity Clouds},
  journal = {The Astrophysical Journal Supplement Series},
  year    = {2001},
  volume  = {136},
  pages   = {463--535},
  doi     = {10.1086/321783},
  eprint  = {astro-ph/0102147},
  archivePrefix = {arXiv}
}

@article{Bregman1980GalacticFountain,
  author  = {Bregman, Joel N.},
  title   = {The Galactic Fountain of High-Velocity Clouds},
  journal = {The Astrophysical Journal},
  year    = {1980},
  volume  = {236},
  pages   = {577--591},
  doi     = {10.1086/157776}
}

@article{FraternaliBinney2008Accretion,
  author  = {Fraternali, Filippo and Binney, James J.},
  title   = {Accretion of Gas on to Nearby Spiral Galaxies},
  journal = {Monthly Notices of the Royal Astronomical Society},
  year    = {2008},
  volume  = {386},
  pages   = {935--944},
  doi     = {10.1111/j.1365-2966.2008.13071.x},
  eprint  = {0802.0496},
  archivePrefix = {arXiv}
}

@article{TothOstriker1992DiskInfall,
  author  = {T{\'o}th, G{\'a}bor and Ostriker, Jeremiah P.},
  title   = {Galactic Disks, Infall, and the Global Value of Omega},
  journal = {The Astrophysical Journal},
  year    = {1992},
  volume  = {389},
  pages   = {5--26},
  doi     = {10.1086/171185}
}
\bibliographystyle{aasjournalv7}



\end{document}